\newif\ifsubmode
\submodefalse

\newif\ifprintfig
\printfigtrue

\ifsubmode
  \documentclass{aastex}
  \usepackage{epsf}

  \cpright{}{}
\else
  \documentclass[11pt,preprint]{aastex}
  \usepackage{epsf}
  \slugcomment{{\it Accepted for AJ \today}}
\fi

\shortauthors{Verdoes Kleijn et al.}
\shorttitle{The black hole in NGC 4335}

\newcommand{\etal}{{et al.~}}

\newcommand{\lta}{\lesssim}
\newcommand{\gta}{\gtrsim}

\newcommand{\grad}{^{\circ}}
\newcommand{\kms}{\>{\rm km}\,{\rm s}^{-1}}
\newcommand{\ergscm}{\>{\rm erg}\,{\rm s}^{-1}\,{\rm cm}^{-2}}
\newcommand{\pc}{\>{\rm pc}}
\newcommand{\Mpc}{\>{\rm Mpc}}
\newcommand{\Msun}{\>{\rm M_{\odot}}}
\newcommand{\Lsun}{\>{\rm L_{\odot}}}
\renewcommand{\deg}{^{\circ}}

\newcommand{\Mbh}{M_{\bullet}}
\newcommand{\Rt}{R_{\rm t}}
\newcommand{\Vc}{V_{\rm c}}

\newcommand{\HalphaNII}{H$\alpha$+[\ion{N}{2}]}
\newcommand{\Halpha}{H$\alpha$}
\newcommand{\NII}{[\ion{N}{2}]}
\newcommand{\SII}{[\ion{S}{2}]}

\begin{document}

\title{Gas Kinematics and the Black Hole Mass at the Center of the Radio Galaxy NGC 4335\altaffilmark{1}}

\author{%
Gijs A.~Verdoes Kleijn,\altaffilmark{2,3}
Roeland P.~van der Marel,\altaffilmark{3}\\
P.~Tim de Zeeuw,\altaffilmark{2}
Jacob Noel-Storr \altaffilmark{4}
and Stefi A.~Baum\altaffilmark{3}
}

\altaffiltext{1}{Based on observations with the NASA/ESA Hubble Space
       Telescope obtained at the Space Telescope Science Institute, which is
       operated by the Association of Universities for Research in Astronomy,
       Incorporated, under NASA contract NAS5-26555.}

\altaffiltext{2}{Sterrewacht Leiden, Postbus 9513, 2300 RA Leiden,
The Netherlands.}

\altaffiltext{3}{Space Telescope Science Institute, 3700 San Martin Drive,
Baltimore, MD 21218.}

\altaffiltext{4}{Columbia University, Department of Astronomy,
538 West 120th Street, New York, NY 10027.}


\ifsubmode\else
\clearpage\fi


\ifsubmode\else
\baselineskip=14pt
\fi


\begin{abstract}
  
  We investigate the kinematics of the central gas disk of the
  radio-loud elliptical galaxy NGC 4335, derived from HST/STIS
  long-slit spectroscopic observations of {\HalphaNII} along 3
  parallel slit positions. The observed mean velocities are consistent
  with a rotating thin disk. We model the gas disk in the customary
  way, taking into account the combined potential of the galaxy and a
  putative black hole with mass $\Mbh$, as well as the influence on
  the observed kinematics of the point spread function and finite slit
  width. This sets a 3$\sigma$ upper limit of $10^8 \Msun$ on $\Mbh$.
  The velocity dispersion at $r \lta 0.5''$ is in excess of that
  predicted by the thin rotating disk model. This does not invalidate
  the model, if the excess dispersion is caused by localized turbulent
  motion in addition to bulk circular rotation. However, if instead
  the dispersion is caused by the BH potential then the thin disk
  model provides an underestimate of $\Mbh$. A BH mass $\Mbh \sim 6
  \times 10^8 \Msun$ is inferred by modeling the central gas
  dispersion as due to an isotropic spherical distribution of
  collisionless gas cloudlets. The stellar kinematics for NGC 4335 are
  derived from a ground-based (WHT/ISIS) long-slit observation along
  the galaxy major axis. A two-integral model of the stellar dynamics
  yields $\Mbh \gta 3 \times 10^9 \Msun$.  However, there is reason to
  believe that this model overestimates $\Mbh$.
  
  Reported correlations between black hole mass and inner stellar
  velocity dispersion $\sigma$ predict $\Mbh$ to be $\geq 5.4 \times
  10^8 \Msun$ in NGC 4335. If our standard thin disk modeling of the
  gas kinematics is valid, then NGC 4335 has an unusually low $\Mbh$
  for its velocity dispersion. If, on the other hand, this approach is
  flawed, and provides an underestimate of $\Mbh$, then black hole
  masses for other galaxies derived from HST gas kinematics with the
  same assumptions should be treated with caution.
  
  In general, a precise determination of the $\Mbh - \sigma$ relation
  and its scatter will benefit from (i) joint measurements of $\Mbh$
  from gas and stellar kinematics in the same galaxies and (ii) a
  better understanding of the physical origin of the excess velocity
  dispersion commonly observed in nuclear gas disks of elliptical
  galaxies.

\end{abstract}



\clearpage


\section{Introduction}
\label{s:intro}

The Hubble Space Telescope has made it possible to measure the masses
of black holes (BHs) in the centers of many nearby active and
quiescent galaxies using stellar and/or gaseous kinematics (for
reviews, see e.g., Richstone \etal 1998; van der Marel 1999; Ho 1999;
de Zeeuw 2001; Kormendy \& Gebhardt 2001). To date, BH masses have
been measured in about 40 galaxies, both spirals and ellipticals, and
this number continues to increase. The BH masses correlate loosely
with host spheroid luminosity (Kormendy \& Richstone 1995) and more
tightly with inner stellar velocity dispersion (Gebhardt \etal 2000; Ferrarese
\& Merritt 2000).

Central emission-line gas is detected in virtually all nearby radio
galaxies, defined here as galaxies which harbor kpc-scale radio-jets.
The equivalent widths of the gas emission lines are generally much
larger than those of the absorption lines in the integrated stellar
light, so that measurement of the kinematics of nuclear emission-line
gas is an efficient way to determine the central gravitational
potential and BH mass (e.g., Harms \etal 1994; Ferrarese, Ford \&
Jaffe 1996; Macchetto \etal 1997; van der Marel \& van den Bosch 1998;
Ferrarese \& Ford 1999; Verdoes Kleijn \etal 2000, hereafter VK00;
Sarzi \etal 2001; Barth \etal 2001). A drawback of this approach is
that the gas kinematics might be affected by non-gravitational
motions. Nevertheless, using the gas kinematics to determine accurate
BH masses is particularly interesting for radio galaxies because
determining the BH mass and the properties of the gas disk can advance
our understanding of radio-jet formation and evolution. In turn, the
radio-jet offers an extra diagnostic of the BH accretion and immediate
BH surroundings.  Current questions include: what is the lower-limit
to BH masses that can form kpc-scale jets? Is this lower-limit higher
or lower than typical BH masses in spirals which (almost) never show
kpc-scale jets? Is there a correlation between BH mass and jet
properties, such as total power or jet velocity?  For these and other
reasons we are performing a systematic study of a complete sample of
nearby radio galaxies with Fanaroff \& Riley (1974) Type I radio jets,
the `UGC FR-I sample' (Verdoes Kleijn \etal 1999, hereafter VK99; Xu
\etal 2000) using observations at multiple wavelengths. In particular,
we have performed a STIS spectroscopic survey of the inner gas
distributions to measure the kinematics and the physical state of the
gas (Noel-Storr \etal 2002).

Here we concentrate on one galaxy from our sample, NGC 4335, which is
a relatively unknown isolated giant elliptical ($M_B=-20.7^m$; Paturel
\etal 1997) at 66 Mpc. The gas appears embedded in a well-defined dust
disk (diameter $\sim 750 \pc$) and gas kinematics can be traced
sufficiently far out along the three slits to allow detailed gas
dynamical modeling. In addition we perform stellar dynamical modeling
for NGC 4335 using the stellar kinematics derived from a WHT/ISIS
long-slit observation.

The paper layout is as follows. Sections~\ref{s:wfpc2} and
\ref{s:spec} present HST/WFPC2 broad- and narrowband imaging, HST/STIS
gas emission-line spectroscopy and ground-based WHT/ISIS stellar
absorption-line spectroscopy, including the basic data reduction and
derivation of the gaseous and stellar kinematics.
Section~\ref{s:modelH} describes the modeling of the gas disk flux
distribution, the derivation of the stellar mass distribution and the
fits to the observed gaseous kinematics to estimate the BH mass.
Section~\ref{s:starkin} describes two-integral modeling of the WHT
stellar kinematics to determine the stellar mass-to-light ratio and to
constrain the BH mass independently.  Section~\ref{s:discon} discusses
the implications of the BH mass measurements in NGC 4335 for our
understanding of BH demography, and for the techniques used to measure
black hole masses.

We adopt $H_0 = 70 \kms \Mpc^{-1}$ throughout this paper. This does
not directly influence the data-model comparison for any of our
models, but does set the length, mass and luminosity scales of the
models in physical units. Specifically, distances, lengths and masses
scale as $H_0^{-1}$, while mass-to-light ratios scale as $H_0$.

\section{Imaging}
\label{s:wfpc2}

\subsection{WFPC2 Setup and Data Reduction}
\label{s:wfpc2_red}

We imaged NGC 4335 in the context of HST program GO-6673. We used the
WFPC2 instrument (described in, e.g., Trauger \etal 1994; Biretta
\etal 1996) on March 15, 1997 to obtain images of NGC 4335 in the
F555W ($V$) and F814W ($I$) filters and linear ramp filter FR680N
which includes {\HalphaNII}. The observing log is presented in
Table~\ref{t:WFPC2}.  The `Linear Ramp Filters' (LRFs) of the WFPC2
have a central wavelength that varies as a function of position on the
detector. The LRF FR680N image (WF CCD) was used as `on-band' image,
with the galaxy position chosen so as to center the filter
transmission on the {\HalphaNII} emission lines. A combination of the
broad-band images F555W and F814W (PC CCDs) was chosen as `off-band'
image and covers primarily stellar continuum. This combination of
filters ensures that the effect of dust extinction is the same in the
on- and off-band image. The final {\HalphaNII} emission image was
obtained by subtracting the stellar continuum, as determined by the
off-band image, from the on-band image. The quality of the
{\HalphaNII} image is set by the Poisson noise of input images and the
uncertainty in the alignment and scaling of the on-band image and
off-band image. The stellar surface brightness is steeply rising at
the center of NGC 4335. The resulting emission image depends quite
sensitively on precise sub-pixel alignment and scaling. This
introduces a $\sim 15\%$ flux error which is estimated by varying the
alignment and scaling within a reasonable range.  The $V$-band and
{\HalphaNII} images are shown in Figure~\ref{f:images}. Extended
{\HalphaNII} emission is present with a roughly elliptical morphology,
elongated in the same direction as the dust disk. As described in
Section~\ref{s:flux_gas}, the emission-line gas is consistent with
being in a thin disk, but the uncertainties involved in constructing
the image are too large to prove this conclusively. We cannot rule out
a more spheroidal distribution, especially towards the nucleus.

These data were presented previously in VK99, who also give a more
detailed description of the data reduction.

\subsection{The Dust Disk}
\label{s:wfpc2_ana}

The central dust detected in NGC 4335 extends to more than 2 kpc from
the nucleus. In the outer regions `arms' of dust gradually align with
the well-defined central dust disk which appears as roughly half an
ellipse on the west side of the nucleus (see Fig.~\ref{f:images}).
This is commonly seen in galaxies with inclined nuclear dust disks
(e.g., NGC 7052, van der Marel \& van den Bosch 1998; NGC 6251,
Ferrarese \& Ford 1999; 3C 449, Martel \etal 1999). On the east side
no dust obscuration is evident in either the $V$- or $I$-band image
and its $V-I$ color is typical for dust-free giant elliptical
galaxies. We conclude that the eastern half of the dust disk produces
negligible obscuration in the optical.

We fitted by eye an elliptical contour to the outline of the dust disk
in the $V$-band image, and measured the position angle of the major
axis of the dust disk to be PA=$156\deg$. This is well-aligned with
the galaxy major axis just outside the dust disk (cf.\
Fig.~\ref{f:images} and VK99). If we assume that the dust disk is thin
and intrinsically circular, then the inclination is $i=66^{\circ} \pm
7^{\circ}$ (VK99).

The total mass of the kpc-scale dust distribution estimated using the
observed color excess is $\sim 5.8 \times 10^5 \Msun$ if the dust
resides in the mid-plane of the galaxy (VK99).  Following the
prescription by Goudfrooij \& de Jong (1994), the IRAS $60\mu$ and
$100\mu$ flux measurements for NGC 4335 from Knapp \etal (1989)
indicate a dust mass $2.2 \times 10^6 \Msun$. This is not inconsistent
with the aforementioned result, given that the IRAS measurements cover
a much larger area than that occupied by the nuclear dust disk studied
with HST. Furthermore, dust mass estimates using color excess are not
sensitive to a smoothly distributed dust component, if present (cf.\ 
e.g., Tran \etal 2001).

\section{Spectroscopy}
\label{s:spec}

\subsection{HST/STIS Gas Emission-line Spectroscopy}
\label{s:spec_red}

In the context of GO-program 8236, we used STIS (see
Kimble~\etal 1998) on September 22, 1999 to obtain long slit spectra
of NGC 4335 during two orbits of the HST telescope. We obtained spectra
along three adjacent slit positions using a $0.2''$ wide slit
(52X0.2). The lay-out of the slits is shown in Figure~\ref{f:images}.
The G750M grating was used in combination with two pixel spatial and
spectral on-chip rebinning (in order to obtain enough signal to noise while
retaining sufficient spatial and spectral resolution) yielding spectra
covering the wavelength range from 6436 {\AA} to 7100 {\AA} over 511
pixels. Wavelength calibration spectra of the internal arc lamp were
obtained at the beginning of the first orbit and at the conclusion of
the second orbit.

The slits were aligned along PA=$-32.44\deg$. This is almost parallel
($\Delta$PA=$-8.44\deg$) to the galaxy major axis (PA=$156\deg$).
We denote the central slit by C, and the adjacent slits on the east-
and west-side by E and W, respectively. A log of the observations is
provided in Table~\ref{t:STISsetup}. Target acquisition uncertainties
and other possible systematic effects could cause the slit positions
on the galaxy to differ slightly from those commanded to the
telescope. We determined the actual slit positions from the data
themselves, using the STIS continuum and emission-line fluxes and the
WFPC2 images. We compared the continuum counts as function of position
along the slit for the three STIS slits with the counts in the WFPC2
$V$ and $I$ images at these positions. The STIS continuum was
determined blue-ward of {\HalphaNII} and red-ward of {\SII}
respectively. This analysis indicates that the positioning errors for
the STIS slits with respect to the WFPC2 image are less than $0.05''$
in both the direction along and perpendicular to
the slit. Another estimate of the positioning of the
slits with respect to the galaxy nucleus was obtained by interpolation
of the central few {\HalphaNII} fluxes assuming the maximum flux
occurs at the galaxy nucleus. This analysis implies that the central
slit was properly centered on the nucleus to an accuracy of
$0.01''$. The small slews performed by the telescope between
observations to dither the slits and observe the adjacent slits have a
nominal positioning error of $\leq 0.003''$.

Most of the necessary data reduction steps were performed by the
HST/STIS calibration pipeline (CALSTIS version 2.3), including
flat-fielding, hot pixel and cosmic-ray removal, absolute sensitivity
calibration and wavelength calibration which takes into account the
Earth's motion. To facilitate hot pixel and cosmic-ray removal two
exposures were taken at each slit position dithered by 2 pixels in the
direction along the slit. The uncertainty in the relative
wavelength accuracy is $\sim\!0.1${\AA} or
$\sim\!5\kms$ (Diaz-Miller 2001).

\subsection{Gas Kinematics and Fluxes}
\label{s:spec_ana}

The spectra show several emission lines, of which the {\HalphaNII}
complex at 6548, 6563, 6583 {\AA} has sufficiently high
signal-to-noise ratio $S/N$ for a kinematical analysis. The {\SII}
doublet at 6716 and 6731 {\AA} can be fitted only in the central $\sim
0.2''$ in the central slit and at a lower signal-to-noise ratio. The
results for these few points are consistent with the {\HalphaNII}
kinematics described below.  To quantify the {\HalphaNII} gas
kinematics we fitted the spectra assuming that each emission line is a
Gaussian on top of a flat continuum. This yields for each line the
total flux, the mean velocity $V$ and the velocity dispersion
$\sigma$. We fitted the {\NII} doublet under the assumption that the
individual lines have the same $V$ and $\sigma$ and that the ratio of
their fluxes equals the ratio of their transition probabilities (i.e.,
3; see references in Mendoza 1983). The {\HalphaNII} complex is
influenced by blending of the lines, and we made the additional
assumptions that H$\alpha$ and the {\NII} doublet have the same
kinematics.

We did the Gaussian fitting using software described in van der Marel
\& van den Bosch (1998).  Prior to the emission-line fit the spectra
were rebinned along the slit where necessary to obtain sufficient
$S/N$, and hence sufficiently small errors on the gas kinematics
(Table~\ref{t:gaskin}), to discriminate between the black hole masses
in dynamical models discussed in Section~\ref{s:dyn_mod} and later.
The same gaussian fitting was performed on modeled emission-line
spectra which resulted from the dynamical models. Differences between
the velocity profiles (VP) and the single Gaussian fits are discussed
in Section~\ref{s:dyn_mod}.

The Gaussian fits to the emission lines are shown in
Figure~\ref{f:emisfit}.
The resulting kinematics are shown in Figure~\ref{f:gaskin} and
tabulated in Table~\ref{t:gaskin}. The systemic velocity used in
Figure~\ref{f:gaskin} and Table~\ref{t:gaskin} was estimated from the
HST data themselves, by including it as a free parameter in the
dynamical models described below (see Section~\ref{s:dyn_mod}). This
yields $v_{\rm sys} = 4620 \pm 13 \kms$. This result is consistent
with $v_{\rm sys} = 4615 \pm 22 \kms$ by Huchra \etal (1983) and
$v_{\rm sys} = 4595 \pm 47 \kms$ from the RC3 catalog (de Vaucouleurs
\etal 1991).

The mean-velocity profiles are smooth functions of position and are
indicative of a rotating circular disk of emission gas. The rotating
disk picture is further supported by the fact that the mean velocities
on the minor axis are close to $0\kms$ in all three slits. The
velocity dispersion profiles in all three slits show irregularities.
We verified that this result does not depend sensitively on chosen
region to fit the continuum and masked deviant pixels during the fit.
The velocity dispersion peaks at $\sim 240 \kms$ at the nucleus in the
central slit C. In slit W the dispersion peaks near the nucleus.
However, in slit E the dispersion actually dips at the position
closest to the nucleus.  It seems therefore that the kinematics of NGC
4335 are consistent with a rotating gas disk, but with additional
turbulent motion.

We also performed single Gaussian fits to {\Halpha} and the {\NII}
doublet independently. The two resulting flux profiles have very
similar shapes. The mean velocities for the two sets of emission lines
are very similar as well. The median difference between the mean
velocities of the two components is $\sim 20 \kms$. This is slightly
larger than the typical formal errors given in Table~\ref{t:gaskin}.
These formal errors are derived by propagating the spectral flux
errors. We will assume $20\kms$ as a more realistic error estimate for
the {\HalphaNII} mean velocities. The {\Halpha} and {\NII} velocity
dispersion profiles show very similar behavior, but {\Halpha} has
systematically lower velocity dispersions by about $\sim 20 - 70\kms$.
The differences in dispersion could be caused by the fact that (i) the
{\Halpha} and {\NII} have intrinsically different kinematics or (ii)
the {\Halpha} kinematic measurements are affected by stellar {\Halpha}
absorption.  A similar effect and of the same magnitude is seen for
the gas disk in the S0 galaxy NGC 3245 (Barth \etal 2001).

\subsection{WHT/ISIS Stellar Absorption-line Spectroscopy}

A long-slit stellar absorption spectrum at PA$=148\deg$ (i.e., within
$10\deg$ from the major axis of NGC 4335) was obtained in service mode
with the ISIS (blue arm) spectrograph on the 4.2m William Herschel
Telescope on La Palma on June 13, 2001. A slit-width of $1.0''$ was
used with the R300B grating and the EEV12 CCD detector. The dispersion
is 0.86 {\AA}/pix and the spatial scale is $0.19''$/pix.  The observed
wavelength range was $\sim$3430--6900 {\AA}. The observations included
two 1800s exposures of NGC 4335, a 60s exposure of the B2IVp star
BD+33 2642 to facilitate flux calibration, and standard calibration
exposures. The instrumental velocity dispersion $\sigma_{\rm instr}
\approx 75\kms$ was determined from sky lines and the emission-line
widths in the calibration lamp spectra. Both the star and galaxy
spectra were observed with a seeing of FWHM $\sim$$1.2''$.

The data reduction was performed in IRAF using the CCDPROC (Version Dec93) and TWODSPEC
packages. The basic data reduction steps include bias subtraction,
flat-fielding with dome-flats and cosmic-ray removal.  The contribution
from dark current is negligible. The wavelength calibration was done
with contemporary CuAr lamp spectra taken directly before and after
the observations of NGC 4335. Either 31 or 32 arc-lamp emission lines
were used to map the wavelength as a function of CCD coordinate. The
final residuals per fitted emission line are $\sim 0.3$ {\AA} (or
$18\kms$ at 5000 {\AA}). We subsequently performed the wavelength
calibration and spatial rectification to align the wavelength
direction with the CCD columns in one step.  The flux calibration took
into account telescope throughput and atmospheric extinction. The
Galactic extinction towards NGC 4335 is $A_B=0.063$ (Schlegel \etal
1998) and hence negligible.  As a cross-check for the flux calibration
we verified that the broad-band flux observed through a $1'' \times
1''$ central aperture in the WHT/ISIS spectrum and our previously
observed HST $V$-band image (VK99) agreed to within $\sim 25\%$,
correcting for differences in filter passbands and point spread
function (PSF). This level of agreement is satisfactory given the
uncertainties in WHT/ISIS seeing conditions and extinction
corrections.  Finally, after verifying that the final calibrated
spectra of stars and galaxies did not have any remaining large-scale
patterns, we logarithmically rebinned in the wavelength direction to
perform the kinematical analysis.

\subsection{Stellar Kinematics}
\label{s:stellarkin}

We used the wavelength range $\sim 5200$--5700{\AA} in the NGC 4335
absorption spectrum to derive the stellar kinematics. This range
includes the red-shifted Mg $b$, Fe$_1$ and Fe$_2$ absorption
features.  To obtain the stellar kinematics we used the pixel space
fitting method of van der Marel (1994), which compares the NGC 4335
spectrum to a stellar `template' spectrum. A suitable template star
(spectral type K0V) was obtained during a different observing run with
WHT/ISIS in identical setup. To check for template mismatch we also
used stellar templates (spectral types K1III and K2) described in van
der Marel \& Franx (1993).  These data have an instrumental velocity
dispersion $\sigma_{\rm instr} \sim 60\kms$, i.e., very similar to the
$\sigma_{\rm instr}$ of the NGC 4335 spectrum. The inferred stellar
kinematics agree within the errors. The observed stellar kinematics as
a function of radius are listed in Table~\ref{t:starkin} (and also
shown in Figure~\ref{f:starkin} which is discussed in
Section~\ref{s:starkin}). Stars and gas rotate in the same direction
in the core of NGC 4335.  The galaxy spectrum integrated along the
slit over the central $25''$ yields an aperture velocity dispersion
$\sigma_0=282\pm 15\kms$. The $25''$ length corresponds to two
effective radii as determined by fitting the surface brightness
distribution shown in Figure~\ref{f:sbprof}. The listed formal errors
include photon noise, uncertainties in the continuum fitting and the
variation in derived $\sigma_0$ observed by using different stellar
templates. We are not aware of previous determinations of the central
velocity dispersion for NGC 4335, but note that the central value
$\sigma_0= 300 \pm 60\kms$ inferred from the Faber \& Jackson (1976)
relation agrees well with the observations.

\section{Modeling and Interpretation of the {\HalphaNII} kinematics}
\label{s:modelH}

\subsection{Flux Distribution}
\label{s:flux_gas}

To model the {\HalphaNII} gas kinematics we need a description of the
intrinsic (i.e., deconvolved and de-inclined) emission-line flux
distribution. We use fluxes from both the WFPC2 {\HalphaNII} image and
STIS spectra to constrain the intrinsic distribution.
Figure~\ref{f:fluxcompare} compares the actual flux distribution
observed through the STIS slits with the expected flux distribution
through the STIS slits as inferred from the WFPC2 image. The agreement
for slits C and W is good considering the uncertainties in the
absolute flux calibrations (especially of the LRF filter). For slit E
there is a clear discrepancy, but this does not affect the inferred BH
mass (see Section~\ref{s:bestfit}). We model the (face-on) intrinsic
flux profile as a double exponential,
\begin{equation}
  F(R) = F_1 \exp(-R/R_1) + F_2 \exp(-R/R_2),
\label{e:fluxparam}
\end{equation}
and assume that the disk is infinitesimally thin and viewed at an
inclination $i=66\grad$, (cf.~Section~\ref{s:wfpc2_ana}). The total
flux contributed by each of the two exponential components is $I_i =
2 \pi R_i^2 F_i$ ($i=1,2$), and the overall total flux is $I_{\rm
tot} = I_1 + I_2$.

The best-fitting parameters of the model flux distribution were
determined by a simultaneous fit to the WFPC2 and STIS {\HalphaNII}
fluxes. We use a $0.8''\times0.8''$ square region of the {\HalphaNII}
image centered on the nucleus. The flux errors take into account both
Poissonian errors and a $15\%$ error due to uncertainties in the
alignment and subtraction of the on- and off-band image (see
Section~\ref{s:wfpc2_red}). From the STIS observations we use the
{\HalphaNII} flux of the Gaussian emission-line fits
(Section~\ref{s:spec_ana}). The fluxes from Gaussian fits agree within
the errors with fluxes estimated by directly measuring the spectral
lines. The STIS flux errors take into account readnoise and Poissonian
errors. Typically, the STIS fluxes have much smaller relative errors
than the WFPC2 fluxes.  We perform an iterative fit of the double
exponential to the {\HalphaNII} flux data, taking into account flux
errors and the necessary convolutions with the appropriate PSF, pixel
size and aperture size for each setup. The diamonds in
Figure~\ref{f:fluxfit} show the predictions of the model that best
fits all available data simultaneously. This model has parameters $R_1
= 0.049''$, $R_2 = 0.78''$ and $I_1 / I_{\rm tot} = 0.31$.  The
absolute calibration gives $I_{\rm tot} = 7.7 \times 10^{-13} \ergscm$
for {\HalphaNII}. The lowest panel in Figure~\ref{f:fluxfit} shows the
intrinsic (i.e., deconvolved) flux distribution as a function of
radius.

The thin disk flux model provides a decent fit to the observed STIS
fluxes, apart from a slight under-prediction of the central fluxes in
slit W. The WFPC2 measurements are also reasonably well-fitted,
although the observed flux profile on the major axis suggests a
narrower central flux peak. A second fit was made as well in which we
included only the WFPC2 data. This resulted in a considerably narrower
intrinsic profile (Figure~\ref{f:fluxfit}). However, the two model
emission-flux distributions result in BH masses which agree to within
$\sim 25\%$ as discussed in Section~\ref{s:bestfit}.

\subsection{The Stellar Luminosity Density} 
\label{s:mass_stars}

For the purpose of dynamical modeling we parameterize the
three-dimensional stellar luminosity density $j$. We assume that $j$
is oblate axisymmetric, that the isoluminosity spheroids have constant
flattening $q$ as a function of radius, and that $j$ can be
parameterized as
\begin{equation}
  j(R,z) = j_0 (m/a)^{\alpha} [1+(m/a)^2]^{\beta} , \quad
  m^2 \equiv R^2 + z^2 q^{-2} .
\label{e:lumdendef}
\end{equation}
Here $(R,z)$ are the usual cylindrical coordinates, and $\alpha$,
$\beta$, $a$ and $j_0$ are free parameters. When viewed at inclination
angle $i$, the projected intensity contours are aligned concentric
ellipses with axial ratio $q'$, with $q'^2 \equiv \cos^2 i + q^2
\sin^2 i$. The projected intensity for the luminosity density $j$
is evaluated numerically.

We assume $i=66^{\circ}$ as discussed in Section~\ref{s:wfpc2_ana}. We
take $q'=0.81$ within a radius of $3''$ based on the isophotal
analysis of VK99 (their Figure 2). They determined that the isophotes
are very close to elliptical and that $q'$ varies from $0.81$ to
$0.76$ between major-axis radius $R \sim 3''$ and $R \sim 7''$.
The isophotal major-axis PA=$156\deg$ is nearly constant, varying by
less than $<5\deg$ over the same radius range. A model with constant
$q$ and PA is expected to be adequate in view of the main
uncertainties for the BH mass from our dynamical modeling as discussed
in Section~\ref{s:discon}. For example, we
verified that the variation in ellipticity changes the derived circular
velocities by $<6\%$, which is small compared to our velocity errors.

We determine the mean flux profile over three rows (corresponding to a
width $\sim 0.14''$) along the unobscured eastern galaxy semi-minor
axis (cf.~Section~\ref{s:wfpc2_ana}). We use the $I$-band fluxes as
opposed to $V$-band fluxes because the flux contribution of
emission-lines is negligible in this wavelength region. The observed
surface brightness profile is shown in Figure~\ref{f:sbprof}. The
projected intensity profile of the best-fit model using the fluxes on
the semi-minor axis between $r=0.0''$ and $r=10''$ is shown by the
solid curve. This model has $\alpha=-0.36$, $\beta=-1.02$, $a=0.35''$
and $j_0=1.6 \times 10^2 \Lsun \pc^{-3}$.

The surface mass density of the dust disk, which is proportional to the
optical extinction (cf.\ VK99; and references
therein), does not show a strong increase towards the center but
varies by just about a factor of two over the disk (cf.\ Figure~2 in
VK00). Taking a Galactic value of the gas-to-dust
ratio of 100 (e.g., Bohlin, Savage \& Drake 1978), the total mass of
the disk, which extends over a region $r \sim 2.5''$ is $\sim 5.8 \times
10^7\Msun$. The ratio of enclosed disk mass to stellar mass is at most
$\sim 6 \times 10^{-4}$ in the range $m=[10\pc,1000\pc]$ (i.e.,
$m=[0.03'',3'']$.  Hence we can neglect the gravitational potential
exerted by the gas and dust disk.

There is one complication in estimating the stellar mass profile. The
central $\sim 0.23''$ of the galaxy has a color $V-I \sim 1.35$ on
both the eastern and western side of the galaxy nucleus. This color is
typical of dust-free giant ellipticals.  It implies that the dust
either does not extend all the way to the center (i.e., has a ring
structure) or becomes optically thick towards the center. In the
latter case it would hide approximately half of the stellar mass in
this region as the dust is very close to the center.  We discuss this
scenario in Section~\ref{s:bestfit} and find that it requires an
unusually low stellar mass-to-light ratio to explain the observed
gas~kinematics.

\subsection{Dynamical Models for a Thin Rotating Gas Disk}
\label{s:dyn_mod}

Our thin-disk models for the gas kinematics are similar to those
employed in van der Marel \& van den Bosch (1998) and VK00. The galaxy
model is axisymmetric, with the stellar luminosity density $j(R,z)$
chosen as in Section~\ref{s:mass_stars} to fit the available surface
photometry. The stellar mass density $\rho(R,z)$ follows from the
luminosity density upon the assumption of a constant mass-to-light
ratio $\Upsilon$. $\Upsilon$ is included as a free parameter in the
dynamical model. We assume that the gas is in circular motion in an
infinitesimally thin disk in the equatorial plane of the galaxy, and
has the circularly symmetric flux distribution $F(R)$ given in
Section~\ref{s:flux_gas}. Thus the position angle of the gas disk is
set by the position angle of the galaxy which is known to an accuracy
of $\sim 1\deg$. The inclination $i$ of the gas disk is assumed to be
identical to the dust disk inclination. Below, we will verify PA and
$i$ also from the two-dimensional gas kinematics independently. The
circular velocity $\Vc(R)$ is calculated from the combined
gravitational potential of the stars and a central BH of mass $\Mbh$.
The line-of-sight velocity profile of the gas at position $(x,y)$
on the sky is a Gaussian with mean $\Vc(R) \sin i\frac{x}{R}$ and
dispersion $\sigma_{\rm gas}(R)$, where $R = \sqrt{x^2 + (y/\cos
  i)^2}$ is the radius in the disk. The velocity dispersion of the gas
is assumed to be isotropic, with contributions from thermal and
non-thermal motions: $\sigma_{\rm gas}^2 = \sigma_{\rm th}^2 +
\sigma_{\rm turb}^2$. For practical reasons, we refer to the
non-thermal contribution as `turbulent', although its origin is not
certain (see Section~\ref{s:dispersion}). We parameterize $\sigma_{\rm
  turb}$ through:
\begin{equation}
  \sigma_{\rm turb}(R) = \sigma_{\rm t0} + [\sigma_{\rm t1} \exp(-R/\Rt)] .
\label{e:turbdef}
\end{equation}
This functional form is meant merely to provide a fit to the observed
dispersion and is not based on an underlying physical mechanism for
the turbulence. The temperature of the gas is not an important
parameter: the thermal dispersion for $T \approx 10^4 {\rm K}$ is
$\sigma_{\rm th} \approx 10\kms$, and is negligible with respect to
$\sigma_{\rm turb}$ for all plausible models.

The predicted VP for any given observation is
obtained through flux-weighted convolution of the intrinsic VPs with
the PSF of the observation and the size of the aperture. The STIS PSF
is represented as a sum of 5 Gaussians (see Table~\ref{t:psfstis}). We
ignore the effect of `pixel-bleeding' for the STIS PSF, since this
changes the kinematics predicted by the model only by a few $\kms$
(Barth \etal 2001). The convolutions are described by the
semi-analytical kernels given in Appendix~A of van der Marel (1997),
and were performed numerically using Gauss-Legendre integration. A
Gaussian is fit to each predicted VP for comparison to the observed
$V$ and $\sigma$.
We define a $\chi^2$ quantity that measures the
quality of the fit to the kinematical data, and the best-fitting model
was found by minimizing $\chi^2$ using a `downhill simplex'
minimization routine (Press \etal 1992).

As mentioned, the observed VP and the VP predicted by the dynamical
modeling are fitted with a single Gaussian. Thus we have to verify
that the deviations from a Gaussian are the same in observations and
model. Figure 2 shows that in some spectra, especially near the galaxy
nucleus, the emission lines tend to have smaller peaks and broader
wings than the Gaussian fit. The same trend is indeed seen in the
modeled spectra. We verified that, apart from this difference, there
are no systematic differences as a function of radius between the VP
and the Gaussian fit. Furthermore, the modeled spectra allow us to
determine the difference between the moments of the full VP and the
Gaussian fit qualitatively, also for blended {\HalphaNII} emission
lines, as is the case in many spectra. The representation by single
Gaussian fits turns out to be also quantitatively adequate in the
sense that the differences between true and gaussian $v$ and $\sigma$
of the VPs are typically within the quoted errors for the observed
Gaussian moments.

\subsection{A Thin Disk Fit to the Gas Mean Velocities}
\label{s:bestfit}

We first analyze the {\HalphaNII} gas disk mean velocities and defer
discussion of the velocity dispersions to Section~\ref{s:dispersion}.
We start out with models with no BH in which the gas disk has the
orientation as determined from the photometry. Thus the gas disk is
assumed to have the same inclination $i=66\deg$ and position angle
PA=$156\deg$ as the dust disk and galaxy (see
Section~\ref{s:wfpc2_ana}).  This corresponds to an intrinsic axial
ratio of $q=0.77$ for our spheroid model. The STIS long-slit
observations were performed at a small angle $\theta=-8.44\deg$ to the
major axis (cf.\ Section~\ref{s:spec_red}).

First we vary the $I$-band mass-to-light ratio $\Upsilon_I$ and find
that a model with $\Upsilon_I=3$ in solar units fits the observed gas
kinematics best (see Figure~\ref{f:mbhrange}). We will refer to this
model as the `standard model'. There are some differences between the
model and observations. For slit E, the model predicts a steeper mean
velocity gradient than observed. For slit W, the observed mean
velocities around $x \sim 0.5''$ are systematically larger than
predicted by the model. The total and reduced $\chi^2$ of this model
with one free parameter are $89$ and $3$, respectively. However,
almost $40\%$ of the total $\chi^2$ is contributed by just the two
points at $x=0.406''$ in slit W and at $x=-0.356''$ in slit E. These
two large deviations might be caused by local deviations from the
model, for instance local turbulence. In fact, at the latter location
the velocity dispersion shows an off-nuclear peak. Discarding these
two points, the reduced $\chi^2=2$. Thus given the idealization of the
model, a rotating disk model with localized turbulence is not too
unlikely.

We can estimate $\Upsilon_I$ independently from existing empirical
correlations between $\Upsilon$ and galaxy optical luminosity. For NGC
4335, $m_B=13.5^m$ (Paturel \etal 1997) and we assume $B-V=1.0$,
$V-I=1.3$ and $R-I=0.45$, which is typical for bright ellipticals and
consistent with our WFPC2 observations (cf.\ VK99). We derive
$\Upsilon_I=2.5$ using the relation between $\Upsilon_R$ and absolute
$B$ magnitude determined by van der Marel (1991). The value
$\Upsilon_I=3.5$ is found using the correlation between $\Upsilon_V$
and absolute $V$ magnitude as determined by Magorrian \etal
(1998). The two values are consistent because both correlations show
scatter of a factor $\sim 1.5$ in the observed mass-to-light values at
given host luminosity. The value $\Upsilon_I=3.0$ derived from the gas
kinematics falls nicely in the expected range $2.5 < \Upsilon < 3.5$.

The inclination of the galaxy and its embedded gas disk are assumed to
be similar to the larger scale dust disk. Which $i$ is preferred by
the gas disk kinematics alone? To address this question, we derive the
stellar mass model and intrinsic emission-line flux model assuming
$i=40\deg$ ($q=0.4$) and $i=80\deg$ ($q=0.8$). The profile shape of
our stellar mass model (Section~\ref{s:mass_stars}) does not depend on
inclination but the central stellar mass density changes, i.e., it
increases for decreasing inclination. The changes in the intrinsic
emission-line flux profile are negligible. For a given $\Upsilon_I$,
the increase in central stellar density increases the predicted
velocities, while the decrease in inclination decreases the
line-of-sight component of the predicted velocities. Thus these two
effects compete. As Figure~\ref{f:upsilon} shows, compared to the fit
for $i=66\deg$ the fits are poorer for $i=40\deg$ and $i=80\deg$ and
the best fit $\Upsilon_I \sim 4 - 5$ is higher than expected from the
empirical correlations with host luminosity.

The PA of the gas disk is fixed in the model to be identical to that
of the galaxy major axis, which is determined to within $\sim 1\deg$ by
the WFPC2 isophotal analysis. We have varied the PA$_{\rm d}$ of the
gas disk major axis by $\pm 20\deg$, and find that the circular disk
model fits best for PA$_{\rm d}=152.5\deg$, i.e. $3.5\deg$ smaller
than the PA of the galaxy major axis.  The $\chi^2$ increases by $>40$
for PA$_{\rm d}$ which are $\pm 10\deg$ away from PA$_{\rm
  d}=152.5\deg$.  Thus the PA of the gas disk as preferred by the gas
kinematics agrees well with the PA derived from the WFPC2 stellar isophotes.

In summary, we see that the kinematics of the gas disk support the
disk inclination and PA and the mass-to-light ratio $\Upsilon_I$
implied by independent methods. The slope of the central velocity
measurements for each slit are well fitted by this model with no BH.
This a posteriori validates our inclusion of the central velocity
gradient to determine $\Upsilon_I$ because this gradient is evidently
not affected significantly by the central BH mass. It also implies
that the data provide an upper limit to the BH mass.

We now determine this BH mass upper limit. The models indicate that
only the gradient between the central three points closest in absolute
radial distance from the nucleus (i.e., at $|x| < 0.15$ in slit C) are
significantly affected by a BH with mass $\Mbh \gta 10^8 \Msun$. We will therefore
use these mean velocities to compute the $\chi^2$. Since the model
parameters other than $\Mbh$ are optimized to fit the whole velocity
profile, the $\chi^2$ from the central three mean velocities is
expected to follow approximately a $\chi^2$ probability distribution
with $N_{\rm df}=3-1=2$ degrees of freedom (3 velocity measurements
and 1 free parameter, $\Mbh$) and hence an expectation value
$\langle\chi^2\rangle=2$. In agreement with this the standard
model has $\chi^2=1.8$. Figure~\ref{f:chisquare} shows the increase in $\chi^2$ as a function
of increasing BH mass. From the $\Delta\chi^2$ for the one-parameter
model, a $\Mbh > 10^8 \Msun$ is formally ruled out at the 3 $\sigma$
level.

A more conservative upper-limit can be obtained by assuming the gas
disk is maximally face-on. This will decrease the inferred intrinsic
axis ratio of the galaxy (see Section~\ref{s:mass_stars}). Since no
giant ellipticals flatter than E6 are observed, we take $q=0.4$ as the
minimum acceptable axis ratio for NGC 4335. This implies a galaxy
inclination $i=40\deg$.  Figure~\ref{f:chisquare} shows that a $\Mbh >
1.8 \times 10^8\Msun$ is ruled out at more than the $3\sigma$ level
for $i=40\deg$. Such a model cannot fit the complete velocity curves
as well as the standard model for any $\Upsilon_I$
(Figure~\ref{f:upsilon}). Moreover, the best fit is provided by
$\Upsilon_I \approx 5$ for $i=40\deg$ which is at least $40\%$ larger
than expected, as discussed above.

Section~\ref{s:flux_gas} shows that the WFPC2 flux profile suggests a
narrower flux profile than the STIS data. We performed the gas
dynamical modeling also with the double exponential fit to the WFPC2
data only. This results in a $\sim 25\%$ higher upper-limit to the
inferred BH mass. The other notable discrepancy between the WFPC2 and
STIS fluxes in slit E does not influence the BH mass measurement directly.

In Section~\ref{s:mass_stars} we discussed the possibility that there
might be a central dust distribution inside $r \approx 0.23''$ which
is completely opaque. This will alter the assumed stellar mass
density, because inside $r=0.23''$ approximately half of the galaxy
light will be obscured. We determine the stellar mass density assuming
the intrinsic stellar flux inside $r=0.23''$ is two times the observed
flux and assume that the obscuration by dust is negligible outside
$r=1''$. The fitted model has $\alpha=0.0$, $\beta=-1.4$, $a=0.31''$
and $j_0=6.2 \times 10^2\Lsun \pc^{-3}$. In this case a mass-to-light
ratio $\Upsilon_I \approx 1.5$ is needed to obtain a good fit to the
galaxy-dominated kinematics. This is an unusually low $\Upsilon_I$ and
lies well outside the observed scatter of observed absolute magnitude
vs.~$\Upsilon_I$ correlations (cf.\ Section~\ref{s:bestfit}).
Moreover, the emission-line flux is strongly peaked towards the center
which suggests we have a direct view of the nucleus without
significant obscuration by dust. We conclude that NGC 4335 probably
does not have an opaque central dust distribution. If however it is
present, then our upper limit on the BH mass is an overestimate.

In conclusion, there is good agreement between the disk inclination
$i$ and PA as determined from gas kinematics and dust morphology and
between the mass-to-light ratio $\Upsilon_I$ from gas kinematics and
from independent methods. These agreements form support for the thin
disk model, which indicates $\Mbh < 10^8\Msun$.  The quantities $\Upsilon$, $i$ and
PA are mainly constrained by the gas kinematics at $r>0.25''$, which
is outside the BH radius of influence $r_{\rm BH}$ (with $r_{\rm BH} =
G\Mbh/\sigma^2$, and $\sigma$ the central stellar velocity
dispersion). The upper limit to the BH mass is determined from the
velocity gradient at $r <0.15''$. The inferred BH mass is thus based
on the assumption that the thin disk model holds for $r < 0.15''$. The
next Section considers if this assumption is supported by the gas
velocity dispersions.

\subsection{The Gas Velocity Dispersions}
\label{s:dispersion}

The velocity dispersion is only roughly fitted by a model of the form
(\ref{e:turbdef}) with $\sigma_{\rm t0}=50.0\kms$, $\sigma_{\rm
  t1}=200\kms$ and $R_t=0.2''$ (Figure~\ref{f:turb}). The observed
velocity dispersion varies between $\sigma_{\rm gas} \sim 50\kms$ and
$\sigma_{\rm gas} \sim 250\kms$ (Figure~\ref{f:turb}).  The value of
$\sim 50\kms$ corresponds to the instrumental and thermal line
broadening. In the absence of a BH, the induced line broadening due to
differential rotation over the apertures increases the predicted
dispersion to at most $\sim 70\kms$. The presence of a BH with mass
$10^8 \Msun$ would only affect the central aperture in each slit,
increasing the predicted dispersions to $\sim 100\kms$. Thus for many
apertures, the gas shows a larger velocity dispersion than predicted
by the dynamical model of a thin rotating disk. Also a double Gaussian
fit to each emission-line, to fit a broad and a narrow component of
the emission-line, yields a narrow-line component with a velocity
dispersion in excess of that expected from the dynamical model (e.g.,
$\sigma_{\rm gas} \sim 190 \kms$ at the nucleus). Thus, the `excess' velocity
dispersion cannot be ascribed easily to the presence of a second
kinematic component in addition to the thin rotating disk. An excess
velocity dispersion increasing towards the nucleus, i.e., similar to
seen here, is often observed in other gas disks (e.g., van der Marel
\& van den Bosch 1998; VK00; Barth \etal 2001), with M 87 being a
notable exception (Macchetto \etal 1997). Significant excess
dispersion is only absent at $r \gta 0.5''$. These points are
well-fitted by our model which is strong evidence that (i) the model
of a thin disk in rotation applies at these radii and (ii) the model
parameters that depend sensitively on these points, i.e.,
$\Upsilon_I$, PA, and $i$, are trustworthy regardless of the effects
of excess velocity dispersion observed closer to the nucleus.  What
causes the excess velocity dispersion? Is it gravitational or non-
gravitational in origin?

A gravitational origin would imply that the gas rotates at speeds
lower than the circular velocity.  Only if the excess in velocity
dispersion is modest, $\sigma/v \ll 1$, do approximate formulae for
this effect of `asymmetric drift' (e.g., Binney \& Tremaine 1987)
allow the circular velocity to be derived from the observed mean
velocity (e.g., Barth \etal 2001). This regime does not apply to the
observed kinematics in the central regions if interpreted as
asymmetric drift. Unfortunately, the BH mass depends exclusively on
these nuclear apertures. It is therefore useful to determine the BH
mass in the most extreme case that all dispersion in the central
spectrum is gravitational. One option is that the gas forms a
spheroidal distribution of collisionless cloudlets.  To first
approximation we can model the gas assuming a spherical isotropic
model and solve the Jeans equations in similar fashion as done in
VK00. A black hole mass $\Mbh \sim 6 \times 10^8 \Msun$ is then
required to produce the observed $\sigma=240\kms$. The model is
simplistic and deviations from axisymmetry and isotropy can change the
result by factors of a few (de Bruijne, van der Marel, \& de Zeeuw
1996).  Moreover the model assumes a constant number density to flux
density ratio for the emission-line flux. Different ionization
mechanisms (e.g., central point source or shocks) quite likely produce
quite different relations between number density and emission-line
emissivity as a function of radius. A second option is to assume that
asymmetric drift is present {\sl except} at the very center. The
central velocity dispersion is then due entirely to differential
rotation over the aperture. Also in this case, a $\Mbh \sim 6 \times
10^8\Msun$ is required to produce the central $\sigma=240\kms$ for
$i=66\deg$ and $\Upsilon_I=3$. Thus if the thin disk model breaks down
within the central pixel we might have significantly under-predicted
the BH mass.

The thin disk model could be a good approximation if the excess
velocity dispersion is caused by localized turbulence, which leaves
the gas on average in circular rotation. The highly irregular behavior
of the velocity dispersion points indeed to a non-gravitational
origin. The velocity dispersion of the gas peaks towards the nucleus
in slit C and W, but dips towards the nucleus for slit E.
Interestingly, the location of irregular peaks in the velocity
dispersion profile seems to coincide with deviations of the observed
mean velocities compared to the model, especially in slit E.  The
relative flux of \NII6583 and {\Halpha} can shed light on the
excitation mechanism. Figure~\ref{f:lineratio} shows that the flux
ratio \NII6583 / {\Halpha} roughly peaks towards the nucleus in all
slits. This is qualitatively in agreement with the excess dispersion
being due to turbulent shocks as they tend to enhance the
emission-line ratio (Dopita \etal 1997). Interestingly, the maximum
line ratio (\NII6583 / {\Halpha} $\sim 5$) in fact occurs at the
location of the dip in the velocity dispersion in slit E (and a
possible dip in slit W). No special feature is detected in the WFPC2
image (Figure~\ref{f:images}) at these locations. However, the spine
of the straight radio-jet observed at arcsecond-scale resolution with
the VLA (PA=$79\deg$; Wrobel, Machalski \& Condon 2002; Xu \etal 2000)
passes right through the most conspicuous dip in slit E (between
$x=-0.04''$ and $x=-0.12''$) if it continues straight down to the
sub-arcsecond scales of interest here. Unfortunately, no radio jet
emission is detected at VLBA scales (Xu \etal 2000). Surprisingly, the
velocity gradient in slit E is regular and only slightly shallower
than predicted by our standard model. Thus it seems that the irregular
excess in dispersion does not significantly alter the mean rotational
velocity.

The irregularity of the velocity dispersion profile and its
correspondence to changes in the \NII 6583 / {\Halpha} flux ratio favor a
non-gravitational origin for the excess velocity dispersion.  The
irregularities do not correspond to irregularities in the gas mean
velocities. Instead, the mean velocities suggest normal rotation.  These
observations are consistent with a gas disk for which the bulk of the
gas orbits at the circular velocity but locally posesses additional
kinetic energy in the form of turbulent motions.  Testing the
viability of such a hydrodynamical model is very interesting but
beyond the scope of this paper. This leaves room for an alternative
scenario, in which the excess velocity dispersion is predominantly
gravitational (despite its irregular behavior) implying $\Mbh \sim 6
\times 10^8 \Msun$. The WFPC2 imaging and kinematics cannot rule out
such a spherical gas distribution at the very center (Section~\ref{s:wfpc2_red}).

\section{Stellar Dynamical Models}
\label{s:starkin}

The WHT/ISIS stellar kinematics can place constraints on the
mass-to-light ratio $\Upsilon_I$ and the BH mass independently from
the gas kinematics. We in turn discuss the constraint on $\Upsilon_I$
and BH mass. A fully general stellar dynamical modeling of an
axisymmetric galaxy requires `three-integral' models in which the
phase-space distribution function $f(E,L_z,I)$ depends on the orbital
energy $E$, the angular momentum component parallel to the symmetry
axis, $L_z$, and a third, non-classical integral (e.g., Binney \&
Tremaine 1987). However, the inferred mass-to-light ratio from
ground-based observations is generally insensitive to assumptions
about the structure and dynamics of the galaxy. Models with different
inclination (van der Marel 1991) or anisotropy (van der Marel 1994;
1999) yield the same value of $\Upsilon$ to within $\sim 20$\%.

Thus, to derive $\Upsilon_I$, we constructed axisymmetric two-integral
models for NGC 4335 in which the phase-space distribution function
$f(E,L_z)$ depends only on the two classical integrals of motion. We
use the same galaxy mass model as for the gas kinematics (i.e., $\rho
= \Upsilon_I j$ with $j(R,z)$ as derived in Section~\ref{s:mass_stars})
and we also include a BH mass. The modeling procedure is the same as used
(and described in more detail) by VK00 and similar to applications of
models to large samples by, e.g., van der Marel (1991) and Magorrian
\etal (1998).  Figure~\ref{f:starkin} shows the observed and predicted
kinematics as a function of major axis distance. A value of
$\Upsilon_I \approx 3.5$ in solar units provides a good fit to the
stellar mean velocity and velocity dispersion profiles at $r>2''$.
This is $20\%$ higher than inferred from the gas. Given the $\sim
20\%$ uncertainty for $\Upsilon$, the stellar kinematics out to a
radius of $\sim 5''$ are consistent with the mass-to-light ratio
inferred by the gas velocity profile in the inner $\sim 1''$.

The two-integral models can only reproduce the observed peak in the
stellar velocity dispersion in the inner $\sim 2''$ by invoking a BH
of mass $\Mbh \gta 3 \times 10^9\Msun$. If this BH mass is correct,
the HST gas kinematics are resolving the radius of influence $r_{\rm
  BH} \gta 2.1''$. HOwever, these same gas kinematics firmly rule out such a massive BH. Instead the
gas models predict a BH mass which is at least between 5 and 30
times smaller.  This is very surprising, because the two gas modeling
variants discussed in Section~\ref{s:dispersion} ascribe all nuclear
gas kinetic energy to gravitation. The gas is assumed to be in either
a disk or a spherical cloudlet distribution. These two models are
expected to bracket the true nuclear gas distribution. However, there
are reasons to mistrust the high BH mass from stellar dynamical
modeling. If radial anisotropy in the stellar motions is present, the
BH mass is overestimated. As an example, Magorrian \etal (1998)
determined the BH masses in a sample of 36 galaxies with similar
modeling as performed here (i.e., two-integral modeling with
ground-based kinematics and HST-based photometry). They found that for
the 15 most spherical galaxies, a spherical model with radial
anisotropy remove the need for a BH in all cases (see also van der
Marel 1999). It turns out that this degeneracy can be lifted if
stellar kinematics at HST resolution is available (i.e., resolving the
sphere of influence of the BH) and the complete parameter-space for
axisymmetric models is considered with three-integral modeling. Early-type galaxies do indeed harbor BHs but the masses inferred from
two-integral and three-integral axisymmetric modeling differ
systematically. We compare nine E/S0 galaxies in Magorrian et al
(1998) which also have three-integral modeling (as listed in Kormendy
\& Gebhardt 2001). The average ratio of two-integral BH mass over
three-integral BH mass is 3.7 with a minimum of 0.41 and a maximum of
10. Thus it is not too unexpected that, if we were to have stellar
kinematics at HST resolution as well, three-integral modeling would indicate a BH
mass in better agreement with the value $\Mbh 6.0 \times 10^8 \Msun$ as derived from the gravitational modeling of the gas velocity dispersion. By contrast, the factor $>30$ difference in BH mass between the resolved
thin disk modeling and the stellar dynamical modeling seems a far stretch.
Unfortunately, we cannot expect to obtain a stronger constraint on BH mass with more general axisymmetric three integral modeling with the
current data because we (i) lack stellar kinematics at HST resolution
and hence do not resolve the $r_{\rm BH} \approx 0.4''$ of a $\Mbh=6
\times 10^8 \Msun$ and (ii) have ground-based stellar kinematics only
along the major axis and with reliable measurements of only first and second order
moments.

\section{Discussion and Conclusions}
\label{s:discon}

Before analyzing the implications of the inferred black hole mass we
first discuss the two main differences between our gas disk dynamical
modeling and the modeling used for NGC 3245 presented by Barth \etal
(2001) which constitutes the current state-of-the-art.  First, instead
of the observed emission-line flux distribution, we have used a double
exponential fit to the flux distribution. For NGC 3245 the use of an
exponential fit changes the inferred $\Mbh$ by $30\%$.  Moreover, it
accounted less well for the wiggles in the velocity profile. In our
case the deviations of the exponential model from the observed
photometric and spectroscopic emission-line fluxes are estimated to
have a $\sim 25\%$ effect on BH mass (see Section~\ref{s:bestfit}).
Second, we do not incorporate the velocity shifts due to asymmetric
illumination of the slit by the gas disk in the dispersion direction.
In other galaxies this could be important to derive the BH mass as
they produce surface brightness 'caustics' (Maciejewski \& Binney
2001). These caustics are not observed in NGC 4335. STIS observations
of stars by one of us show that the velocity shift amounts to at most
$\sim 30\kms$ for a $0.2''$ wide slit in the extreme case of a star at
the edge of the slit. Only the (central) apertures in the adjacent
slits have an asymmetric flux gradient in the dispersion direction.
The kinematics from these apertures are not taken into account in
deriving the upper limit to $\Mbh$. Hence we conclude that the
velocity shift has negligible effect on the derived $\Mbh$.

Black hole mass $\Mbh$ correlates in general rather tightly with
observed velocity dispersion $\sigma$ in the inner region of galaxies
(Gebhardt \etal 2000; Ferrarese \& Merritt 2000). Recently, Tremaine
\etal (2002) completed a detailed analysis of the correlation and we
will use their results in what follows. They find a best fit
correlation of the form $\log\Mbh = \alpha +
\beta\log(\sigma/\sigma_0)$ with $\alpha=8.13\pm0.06$ and
$\beta=4.02\pm0.32$ for $\sigma_0=200\kms$. The observed spread in the
correlation indicates that the intrinsic dispersion in $\log\Mbh$ is
0.3 dex (perhaps smaller if observational errors are underestimated).
Measuring the central $\sigma$ in similar fashion as in Tremaine \etal (2002) gives $\sigma=282\kms$ for NGC 4335 (cf.\ 
Section~\ref{s:stellarkin}) which corresponds to a predicted $\Mbh=5.4
\times 10^8 \Msun$. The $3\sigma$ upper-limit $\Mbh < 10^8 \Msun$
from the thin disk modeling of the gas mean velocities
(Section~\ref{s:bestfit}) falls well below the relation, even when
including the reported intrinsic dispersion in $\Mbh$. The residuals
are $-0.73$ dex and $-0.43$ dex for the upper-limit of $\Mbh=1.0
\times 10^8 \Msun$ (best-fit model) and $\Mbh=2.0 \times 10^8 \Msun$
(for a maximally face-on gas disk), respectively (see
Figure~\ref{f:bhsigma}). The residuals are even larger if we use the
best-fit relation as determined by Ferrarese (2002) who infers a
larger $\beta=4.58$.  The $\Mbh \gta 3 \times 10^9\Msun$ derived from
stellar dynamics corresponds to an equally large, but positive residual of
$\gta 0.7$ dex.  Finally, the gravitational modeling of the gas
velocity dispersions yields $\Mbh \sim 6 \times 10^8 \Msun$
corresponding to a residual of $0.05$ dex, well within the reported
intrinsic dispersion of the correlation.

Which BH mass are we to trust?  The analysis strongly supports a thin
rotating gas disk model at $r \gta 0.5''$.  The model provides a
reasonable fit to the gas mean velocities and dispersions, the dust
disk morphology and gas disk kinematics indicate the same inclination
and PA, and the $\Upsilon$ from stellar and gas kinematics is
consistent.  However, there are doubts for the validity of the model
at $r \lta 0.5''$. The WFPC2 and STIS fluxes are consistent with a
thin disk surface brightness profile but the signal-to-noise of the
WFPC2 image is too low to rule out a more spherical distribution.  An
excess velocity dispersion with an irregular profile is observed at $r
\lta 0.5''$. Only an ad hoc explanation of localized random motion
exists for this excess. It is not clear if such quasi-stationary
turbulence in a thin disk is physically viable (e.g., Wada , Meurer \&
Norman 2002; and references therein). If the excess velocity
dispersion is (partly) due to gravitational motion around the BH, then
the thin disk model underestimates the true BH mass. Ascribing all gas
kinetic energy, including the excess velocity dispersion, as
counterbalancing the gravitational potential in a simple manner of
isotropically moving collisionless cloudlets yields a BH mass which
agrees well with the $\Mbh - \sigma$ correlation (Tremaine \etal
2002).  Ascribing the nuclear velocity dispersion to rotational motion
from a spatially {\sl unresolved} disk yields an equally good
agreement (Section~\ref{s:dispersion}).  As discussed in more detail
in Section~\ref{s:starkin}, it is not too unreasonable to assume that
the stellar dynamical modeling overestimates the BH mass by a factor
of $\sim 5$. This would be due to radial anisotropy (often observed in
bright ellitpicals such as NGC 4335) which the two-integral modeling
does not take into account. In conclusion, the gas spheroidal model
infers a BH mass in accordance with the empirical $\Mbh - \sigma$
relation, but remains very simplistic. If doubts about the validity of
the thin disk modeling for $r \lta 0.5''$ were proven true, its
inferred BH mass is expected to be an underestimate, driving the
expected true BH mass to higher values in better agreement with the
$\Mbh - \sigma$ relation. Similarly, the expected corrections for the
stellar dynamical model bring its predicted BH mass in better
agreement with the relation.

What are the implications if in reality NGC 4335 indeed harbors a BH
with $\Mbh \sim 6.0 \times 10^8\Msun$?  Our results then suggest that
gas dynamical modeling assuming thin rotating disks cannot be used to
derive accurate black hole masses, {\sl even when all of the following
  are true}: (i) the gas mean velocities very clearly suggest
rotation; (ii) the surrounding dust disk appears regular; (iii) the
inclination and position angle of the inner gas disk are
well-constrained due to the use of three adjacent slits and these
angles are both indicated by the gas disk kinematics and independent
methods; and (iv) the derived $\Upsilon_I$ from gas and stars agree,
ruling out asymmetric drift at $r \gta 0.5''$. This then would cast
doubt on other $\Mbh$ values determined from gas kinematics using
similar models.  Figure~\ref{f:bhsigma} shows that these measurements
have a large influence on the upper-end of the $\Mbh$ detections and
hence on the slope of the correlation. Moreover, to date all
measurements of $\Mbh$ in nearby radio galaxies are based on gas
kinematics.  However, the fact that the inferred BH masses using this
method follow the best-fit relation more closely than the stellar
dynamical measurements argues against this worry of the validity of
the gas dynamical modeling in general. The $\chi^2$ per degree of
freedom as used in deriving the best-fit relation (cf.\ Equation 3 in
Tremaine \etal 2002) is 0.27 for the gas dynamical measurements while
for the stellar dynamical measurements it is 1.30 (1.17 if one
excludes the Milky Way BH mass measurement which is obtained from a
stellar dynamical model which is completely different from those used
for external galaxies). Moreover, the average residual of the gas
dynamical BH mass measurements is positive, counter to what is
expected if the gas mean velocities systematically underestimate the
circular velocity and hence the inferred BH mass.

Independent determinations of $\Mbh$ in galaxies from gaseous and
stellar kinematics observed at high spatial resolution are crucial to
address the worries about the gas dynamical modeling. This has been
performed for IC 1459 (Verdoes Kleijn \etal 2000; Cappellari \etal
2002). Verdoes Kleijn \etal derive from the gaseous kinematics at 6
FOS pointings a black hole mass ranging from $\Mbh=1.5 \times
10^8\Msun$ (thin disk model) to $6 \times 10^8\Msun$ (isotropic
spheroidal model). However, Cappellari \etal (2002) infer $\Mbh=(2.6
\pm 1.1) \times 10^9\Msun$ based on three-integral modeling of
combined ground-based and HST/STIS stellar kinematics. Unfortunately,
the HST stellar kinematics cannot be measured accurately inside the
sphere of influence for this BH mass.  Cappellari \etal also present a more
complete view of the gas kinematics from a recent STIS long-slit
observation. This indicates that while the data are consistent with
the earlier FOS measurements, and the inferred BH mass is similar
(modeled with independently developed software), the gas mean
velocities are rather perturbed in this particular case. Moreover,
IC1459 has an irregular dust distribution.  Thus, the gas and dust properties in
IC 1459 are quite different from the photometrically and kinematically
well-behaved central disk in NGC 4335. An independent $\Mbh$
measurement based on the stellar kinematics for NGC 4335 at HST resolution, and for other galaxies
with similar gas disk kinematics, would test the validity of the gas
dynamical modeling in well-behaved gas disks as opposed to irregular
gas disks.

It is similarly crucial to understand the origin of the excess
velocity dispersion commonly seen in nuclear gas disks. Presently, it
is not clear under which circumstances the thin gas disks become
locally turbulent (for example, gravitational or MHD instabilities)
and if they remain globally stable (e.g., Wada, Meurer \& Norman 2002;
and references therein). A better idea for the origin of the excess is
needed to improve on the highly idealized collisionless spherical
model discussed in this paper.  In this respect, it is also useful to
determine the ionization mechanism of the emission-line gas in nuclear
disks. If it is shocks we should be wary about the modeling with
unperturbed, infinitely thin rotating disks. Furthermore, the disks of
gas and dust might perhaps interact with ambient hot X-ray gas which
is often present in the centers of bright ellipticals (e.g., Gunn
1979). The ultimate goal for gas dynamical modeling is thus to explain
self-consistently the density, ionization state and dynamical state of
the central gas distributions. This requires high S/N two-dimensional
photometry, kinematics and line ratios to determine the dynamics,
ionization state, density and temperature as a function of disk
radius. This will provide not only accurate BH masses but also a vast
improvement in our understanding of the fueling of BHs.

Kiloparsec-scale radio jets are only seen in early-type galaxies but
never in spirals (with one possible exception known to the authors;
Ledlow \etal 2001). For instance, the UGC FR-I galaxies all have
Hubble types E-S0 (cf.\ VK99). Prior to this study, only BH masses
$\Mbh \geq 2 \times 10^8\Msun$ have been reported in nearby radio
galaxies (i.e., NGC 4261, NGC 4374, M87, NGC 5128, NGC 6251 and NGC
7052, see Tremaine \etal 2002 for references). Interestingly, none of
the seven galaxies with a Hubble type later than S0 in the the sample
compiled by Tremaine \etal (2002) have $\Mbh > 1 \times 10^8\Msun$. We
are aware of only one galaxy with a Hubble type later than S0 and a
$\Mbh > 2 \times 10^8\Msun$, which is the Sombrero galaxy (M104,
Hubble type Sa) with $\Mbh = 1.0 \times 10^9\Msun$, but this
measurement is based on models less general than three-integral
stellar dynamical models.  By contrast, all BH mass upper limits and
detections in a sample of 16 mostly early-type disk galaxies (Hubble
type S0-Sb; Sarzi \etal 2001) are below $2 \times 10^8 \Msun$. These
results suggest that the differences in BH mass might be the
underlying factor for this host preference of radio-jets. However, if
our determination of a $3\sigma$ upper limit of $\Mbh < 10^8\Msun$ on
the BH mass of NGC 4335 is correct, then NGC 4335 would illustrate
that BHs with $\Mbh < 10^8\Msun$ are also capable of producing FR-I
radio jets. This would argue against BH mass being the (only)
parameter underlying the host morphology preference of radio galaxies.


\acknowledgments It is a pleasure to thank Marcella Carollo and Chris
O'Dea for stimulating advice, and to thank Chris Benn and Andy
Longmore for carrying out our WHT service observations and providing
helpful suggestions during the analysis. We would like the referee for
his/her comments which helped to improve the paper significantly.
Support for proposal \#8236 was provided by NASA through a grant from
the Space Telescope Science Institute, which is operated by the
Association of Universities for Research in Astronomy, Inc., under
NASA contract NAS 5-26555. This paper made use of the LEDA database:
{\tt http://leda.univ-lyon1.fr}. G.V.K. is grateful to the Leids
Kerkhoven-Bosscha Foundation for financial support for a visit to
STScI in February 2001.


\clearpage



\ifsubmode\else
\baselineskip=10pt
\fi


\clearpage


\ifsubmode\else
\baselineskip=14pt
\fi


\newcommand{\figcapimages} {{\sl Left:} WFPC2 $V$-band image on the PC
  chip of the central $10''$x$10''$ of NGC 4335. North is up and East
  is to the left. The four thin lines outline the position of the 3
  STIS slits. The slits are drawn to scale. The thick line indicates
  the orientation of the brightest side of the radio jet. An central
  dust disk with a major axis of $\sim 5''$ obscures the west side of
  the nucleus. The disk is surrounded by larger lanes of dust. {\sl
    Right:} WF chip {\HalphaNII} image from VK99. The two images have
  the same scale and orientation.\label{f:images}}

\newcommand{\figcapemisfit}
{{\HalphaNII} emission-lines in the STIS spectra of NGC 4335. The
abscissa is the observed vacuum wavelength in {\AA}, and the ordinate
is the flux in relative units (different per panel). Each panel is labeled with the slit
name and the major axis coordinate of the aperture center in arcsec. The
thin curves are the observed spectra. The thick curves are fits
obtained under the assumption that all emission lines are single
Gaussians, as described in Section~\ref{s:spec_ana}. The mean velocity
$V$ and velocity dispersion $\sigma$ of the Gaussian fits are shown in
Figure~\ref{f:gaskin} and listed in Table~\ref{t:gaskin}.\label{f:emisfit}}

\newcommand{\figcapgaskin}
{Mean velocity (top row) and velocity dispersion (bottom row) of the
ionized gas in NGC 4335, inferred from Gaussian fits to the emission
lines in the STIS spectra (as listed in Table~\ref{t:gaskin} and shown
in Figure~\ref{f:emisfit}). The abscissa in each panel is the major
axis coordinate of the aperture center. Each column corresponds to the
slit position as labeled in the top row panels.\label{f:gaskin}}

\newcommand{\figcapfluxcompare}{A comparison between the emission-line
  fluxes as derived from STIS spectroscopy and WFPC2 imaging. The
  three panels show for each slit the observed {\HalphaNII} flux
  density as a function of major axis position. The open circles
  denote the STIS {\HalphaNII} fluxes as derived from Gaussian fits
  (see Section~\ref{s:spec_ana}). The filled circles indicate the flux
  density through the STIS slits as expected from the WFPC2
  {\HalphaNII} emission image. The WFPC2 flux errors are discussed in
  Section~\ref{s:flux_gas}. There is a good agreement for slits C and
  W, but a discrepancy for slit E. Such differences in assumed
  emission-line flux distribution at $>0.2''$ from the nucleus result
  in a negligible change in inferred BH mass. (cf.\ 
  Section~\ref{s:bestfit}).
  \label{f:fluxcompare}}

\newcommand{\figcapfluxfit}{Top: The observed {\HalphaNII}
  emission-line fluxes and their errors (arbitrary units) in the WFPC2
  emission image.  Each panel plots a $0.0996''$ wide cut parallel to
  the galaxy major axis (i.e., as a function of $x$). The distance
  between the cut and the major axis (i.e., $y$) is listed in each
  panel. The open circles and diamonds denote the modeled fluxes
  predicted by fitting a double exponential to the WFPC2 data only and
  WFPC2 and STIS data simultaneously (see bottom Figure). Middle: The
  observed fluxes and modeled fluxes (i.e., simultaneous fit) for the
  spectra of the E, C and W STIS slits. Bottom: The intrinsic flux
  profiles (solid lines: wider profile is from simultaneous fit to
  STIS and WFPC2) in arbitrary units as a function of distance from
  the galaxy nucleus.  The profiles consist of a double exponential
  (cf.\ Equation~\ref{e:fluxparam}). This plot also shows the modeled
  STIS PSF (dotted line) and WF-chip PSF (dashed line). Differences
  between the two modeled flux distributions amount to a $\sim 25\%$
  difference in inferred BH mass.\label{f:fluxfit}}

\newcommand{\figcapsbprof}
{The stellar surface brightness profile of NGC 4335 in the $I$-band as
a function of radius along the minor axis. The open dots show the
observed profile along the eastern semi-minor axis which is unobscured
by dust (cf.\ Section~\ref{s:mass_stars}). The solid line shows the
best fit model for the surface brightness using the luminosity density
parameterization given in Eq.~(\ref{e:lumdendef}).\label{f:sbprof}}

\newcommand{\figcapmbhrange}
{The observed mean velocities (solid dots with $20\kms$ error bars) of the gas
disk in NGC 4335 as function of major axis radius $x$ for slits E, C
and W. The top row shows the complete data set which extends out to a
major-axis radius of $\sim 1.5''$, while the bottom row zooms in on
the inner $\sim 0.6''$. Overplotted are three dynamical models with a
mass-to-light ratio $\Upsilon=3$ (in solar units) which differ only in
black hole mass: $\Mbh=0.0\Msun$ (open diamonds), $\Mbh=1.0 \times
10^{8} \Msun$ (open triangles) and $\Mbh=5.0
\times 10^8 \Msun$ (open squares). The model predictions are
connected by lines to guide the eye.\label{f:mbhrange}}

\newcommand{\figcapupsilon} 
{The $\chi^2$ of the model fit to the mean velocities as a function of
  $I$-band mass-to-light-ratio $\Upsilon_I$ (solar units) for three
  inclinations of the disk.  Models with inclination $i=40\deg$ are
  denoted by the dashed curve, with $i=66\deg$ by the solid curve and
  $i=80\deg$ by the dotted curve. The most face-on model with
  $i=40\deg$ implies an intrinsic galaxy ellipticity $\epsilon=0.6$.
  Models with $i=66\deg$ provide the best fit to the observed mean
  velocities for an $\Upsilon=3$, which is consistent with stellar
  dynamical modeling of groundbased observations (see
  Section~\ref{s:starkin}). An inclination $66\deg$ is also observed
  for the central dust disk of NGC 4335. The horizontal dashed lines
  indicate the 1, 2 and 3 $\sigma$ level for a two parameter (i.e.,
  $\Upsilon$ and $i$) $\chi^2$ model. Models with $i=80\deg$ and
  $i=40\deg$ are formally ruled out at the 2 and 3 $\sigma$ level and
  moreover have a best-fitting $\Upsilon_I \approx 4$ and $\Upsilon_I
  \approx 5$, respectively, which is higher than expected (see
  Section~\ref{s:bestfit}).\label{f:upsilon}}

\newcommand{\figcapchisquare} 
{The $\chi^2$ from the central three {\HalphaNII} mean velocities in
  the central slit as a function of BH mass. Only these mean
  velocities are significantly affected by the BH gravitational field.
  We have assumed mean velocity errors $\Delta v = 20\kms$ (cf.\ 
  Section~\ref{s:spec_ana}). The model parameters other than BH mass
  are fixed to the values that best fit the complete mean velocity
  profile. For the standard model, indicated by the solid dots and
  line this is a mass-to-light ratio $\Upsilon_I=3.0$ and disk
  inclination $i=66\deg$. The horizontal dotted lines indicate the 1,
  2 and 3 $\sigma$ confidence intervals for this $\chi^2$ probability
  distribution for one free parameter.  A black hole mass $\Mbh >
  10^8\Msun$ is excluded at more than the $3\sigma$ level. Also
  plotted are the more conservative upper limits for a model with a
  minimally inclined gas disk of $i=40\deg$, for which  $\Mbh \gta 2 \times 10^8
  \Msun$ is ruled out at the $3\sigma$ level (dashed line, open dots).
  However, this inclination does not agree with the dust disk
  inclination and the best fit $\Upsilon \approx 4.5$ is formally
  ruled out at more than the $3\sigma$ level (cf.\ 
  Figure~\ref{f:upsilon}). See Section~\ref{s:bestfit} for a more
  detailed discussion.\label{f:chisquare}}

\newcommand{\figcapturb}
{The observed velocity dispersions (solid dots with error
bars) of the gas disk in NGC 4335 as a function of major axis
coordinate $x$ for slits E, C and W. Overplotted (squares) is the
standard dynamical model with a black hole $\Mbh=0.0 \Msun$ and a
turbulent velocity dispersion defined by $\sigma_0=50\kms$,
$\sigma_1=200\kms$ and $\Rt=0.2''$ (cf.\
Equation~\ref{e:turbdef}). The model predictions are connected by
lines to guide the eye. Section~\ref{s:dispersion} discusses the
possible origins of this turbulent dispersion.\label{f:turb}}

\newcommand{\figcaplineratio}
{From left to right: the observed flux ratio \NII6583 / {\Halpha} for
slit E, C and W, respectively, as a function of gas disk major axis
coordinate $x$. The peak in slit E spatially coincides with the dip in
$\sigma$ (see Figure~\ref{f:turb}). The projected radio jet crosses
this region as well. This is consistent with a scenario in which the
jet interacts with the gas and induces shocks (see
Section~\ref{s:dispersion}).\label{f:lineratio}}

\newcommand{\figcapstarkin} {Mean stellar velocity (top panel),
  velocity dispersion (middle panel) and root mean square velocities
  ($v_{\rm RMS} \equiv [V^2 + \sigma^2]^{1/2}$; bottom panel) along
  the major axis for NGC 4335. The curves are predictions of
  $f(E,L_z)$ isotropic rotator models with central BH masses of
  $\Mbh=0, 1.5, 3$ and $6 \times 10^9 \Msun$, respectively, and I-band
  mass-to-light ratio $\Upsilon_I=3.5$ in solar units. This
  mass-to-light ratio provides a good fit to the stellar velocities
  for all models and is not inconsistent with, but $20\%$ higher than
  inferred from the gaseous kinematics (cf.\ Section~\ref{s:starkin}
  and Figure~\ref{f:upsilon}). The two-integral modeling without a BH
  does not give a good fit to the ground-based stellar velocity
  dispersions for $r< 2''$ and suggests instead a BH mass $\Mbh \sim 3
  \times 10^9\Msun$. This is at least an order of magnitude higher
  than inferred from the gas. As discussed in Section~\ref{s:starkin},
  two-integral modeling is generally sufficient for an estimate of the
  mass-to-light ratio $\Upsilon$ from ground-based data.  However, a
  robust estimate for the BH mass requires both stellar kinematical
  observations at higher spatial resolution and more general (i.e.,
  three-integral or triaxial) stellar dynamical
  models.\label{f:starkin}}

\newcommand{\figcapbhsigma} {Measurements of BH mass $M_{\rm BH}$ as a
  function of the luminosity-weighted velocity dispersion $\sigma$
  within an effective radius for the sample of galaxies compiled by
  Tremaine \etal (2002) and their linear fit to these data. The BH
  detections for the entire sample are based on dynamical modeling of
  observed kinematics from stars (circles), ionized gas (triangles) or
  masers (squares) in both spirals and early-type galaxies. The solid
  symbols denote galaxies which have kpc-scale radio jets. Three BH
  mass measurements for NGC 4335 are plotted as well. The $3\sigma$
  upper limit of $\Mbh < 10^8 \Msun$ from the thin disk modeling of
  the mean velocities, the $\Mbh \sim 6 \times 10^8\Msun$ from the gas
  collisionless cloudlet modeling of the velocity dispersions and the $\Mbh \sim 3
  \times 10^9 \Msun$ from the stellar dynamical axisymmetric
  two-integral modeling. The validity for these models is discussed in
  summary in Section~\ref{s:discon}. The solid line depicts the
  best-fit relation which predicts $\Mbh=5.4 \times 10^8 \Msun$ for
  the observed $\sigma$. The dashed lines depict the estimated $0.3$
  dex intrinsic dispersion in black hole mass (Tremaine \etal
  2002).\label{f:bhsigma}}


\ifsubmode
\figcaption{\figcapimages}
\figcaption{\figcapemisfit}
\figcaption{\figcapgaskin}
\figcaption{\figcapfluxcompare}
\figcaption{\figcapfluxfit}
\figcaption{\figcapsbprof}
\figcaption{\figcapmbhrange}
\figcaption{\figcapupsilon}
\figcaption{\figcapchisquare}
\figcaption{\figcapturb}
\figcaption{\figcaplineratio}
\figcaption{\figcapstarkin}
\figcaption{\figcapbhsigma}

\clearpage
\else\printfigtrue\fi

\ifprintfig


\clearpage
\begin{figure}
\plottwo{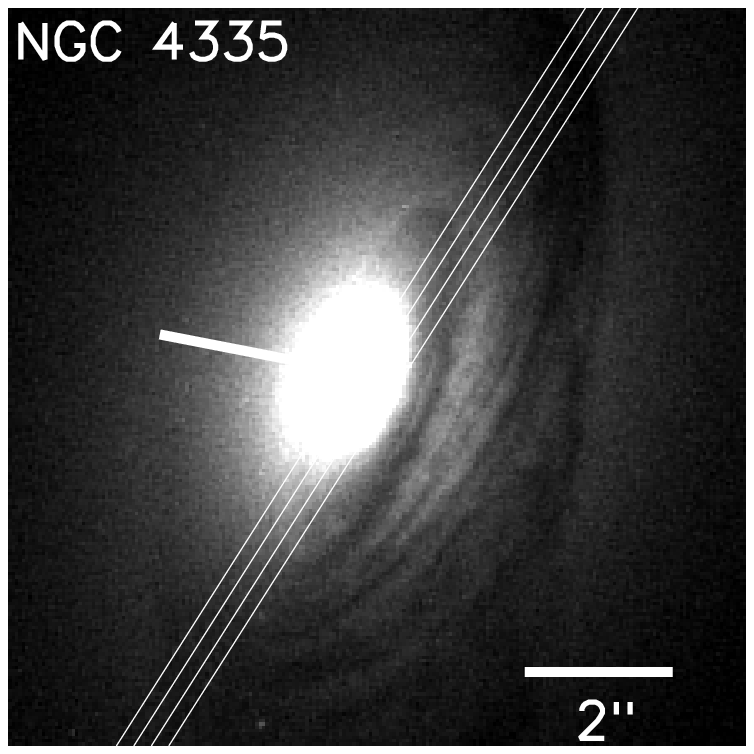}{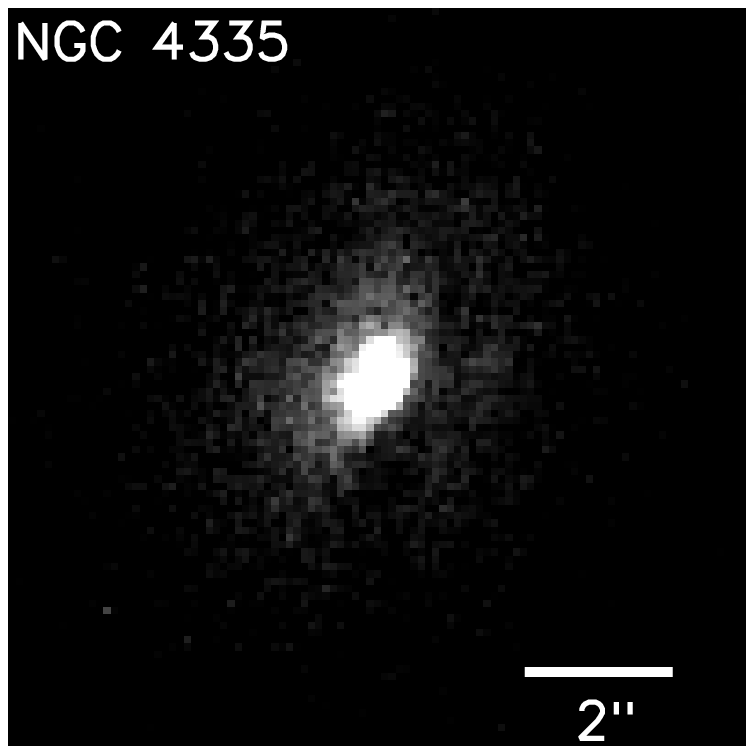}
\ifsubmode
\vskip3.0truecm
\setcounter{figure}{0}
\addtocounter{figure}{1}
\centerline{Figure~\thefigure}
\else\figcaption{\figcapimages}\fi
\end{figure}


\clearpage
\begin{figure}
\centerline{\epsfbox{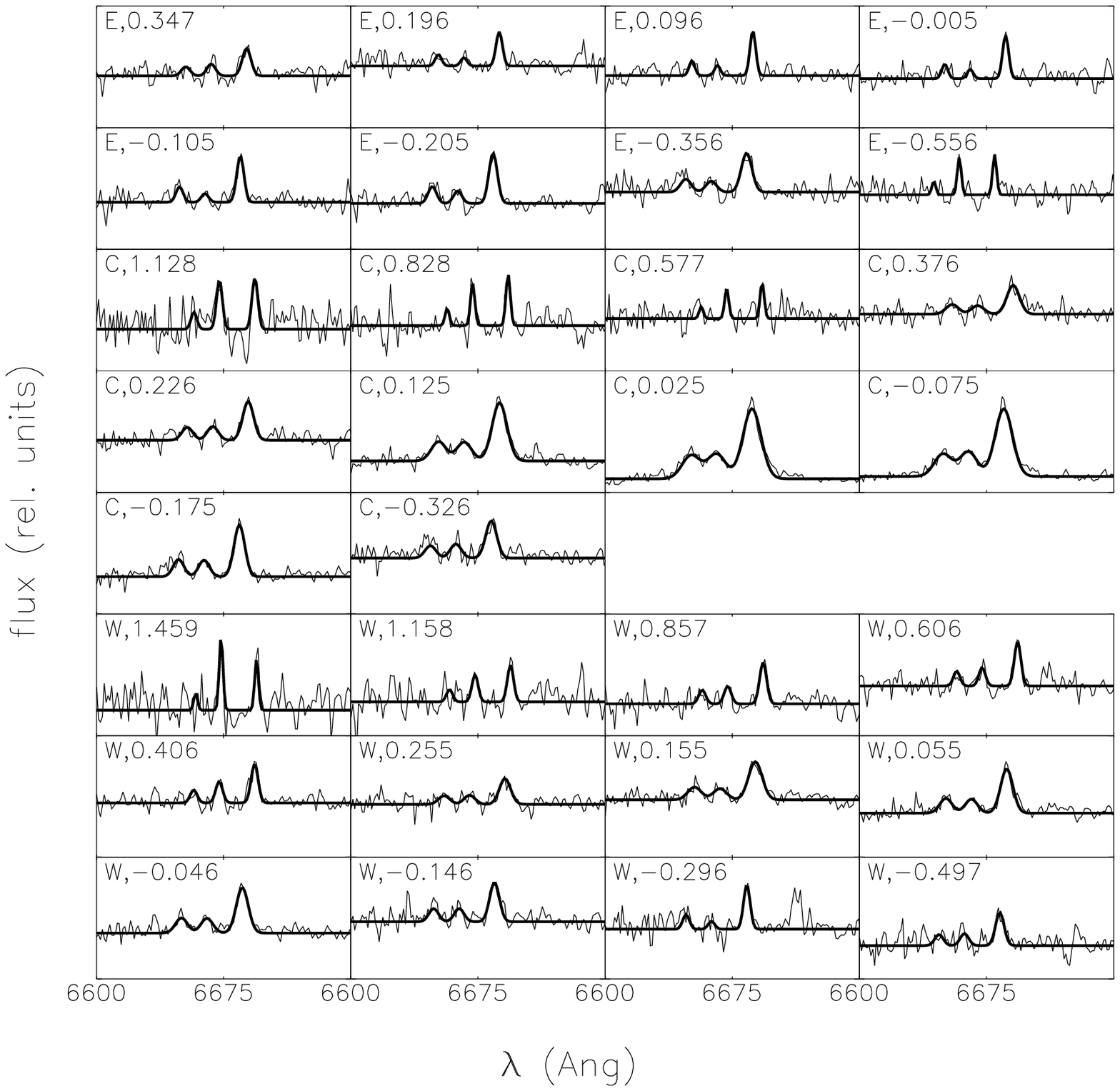}}
\ifsubmode
\vskip3.0truecm
\addtocounter{figure}{1}
\centerline{Figure~\thefigure}
\else\figcaption{\figcapemisfit}\fi
\end{figure}


\clearpage
\begin{figure}
\centerline{\epsfbox{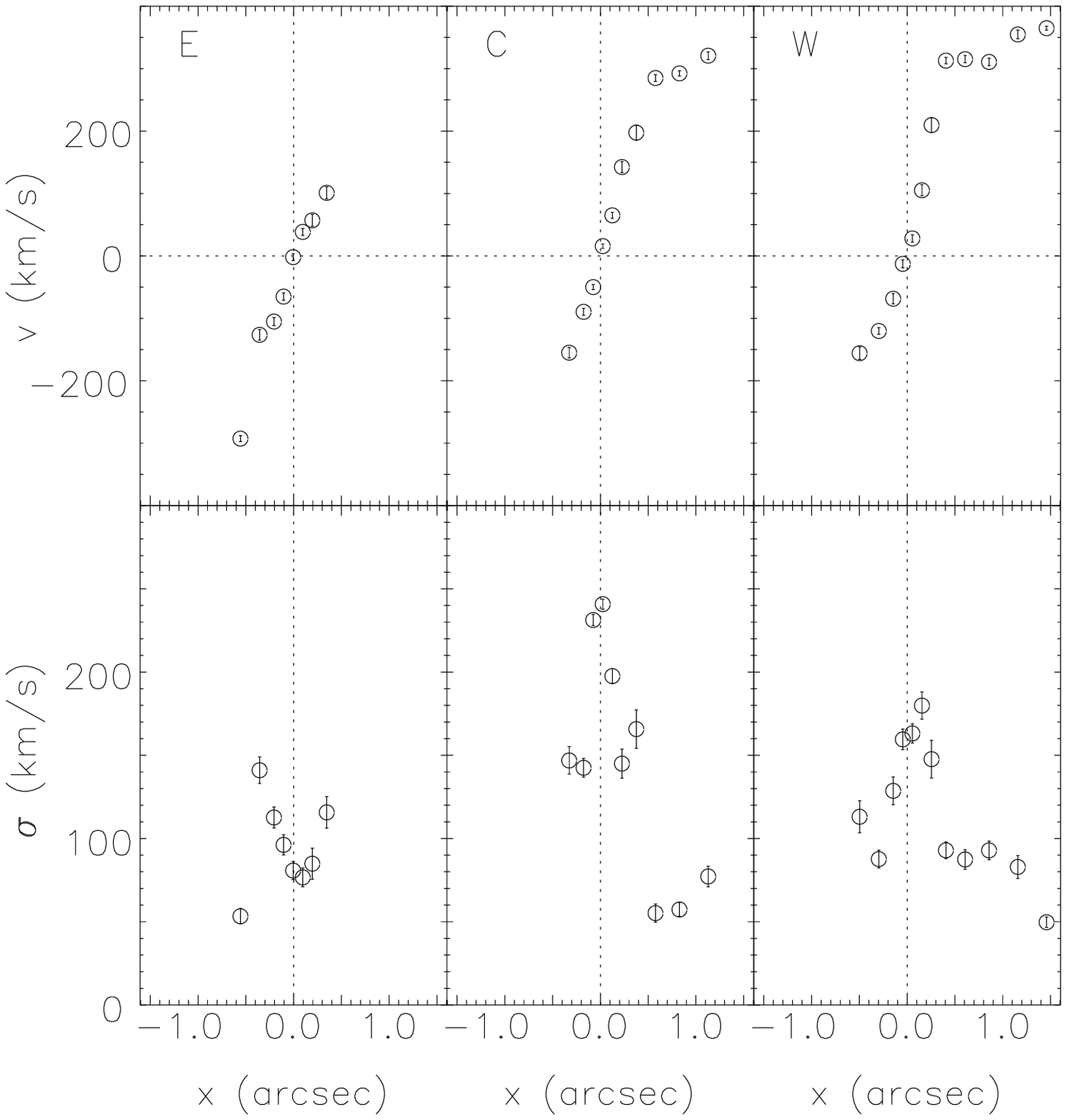}}
\ifsubmode
\vskip3.0truecm
\addtocounter{figure}{1}
\centerline{Figure~\thefigure}
\else\figcaption{\figcapgaskin}\fi
\end{figure}


\clearpage
\begin{figure}
\epsfxsize=0.9\hsize
\centerline{\epsfbox{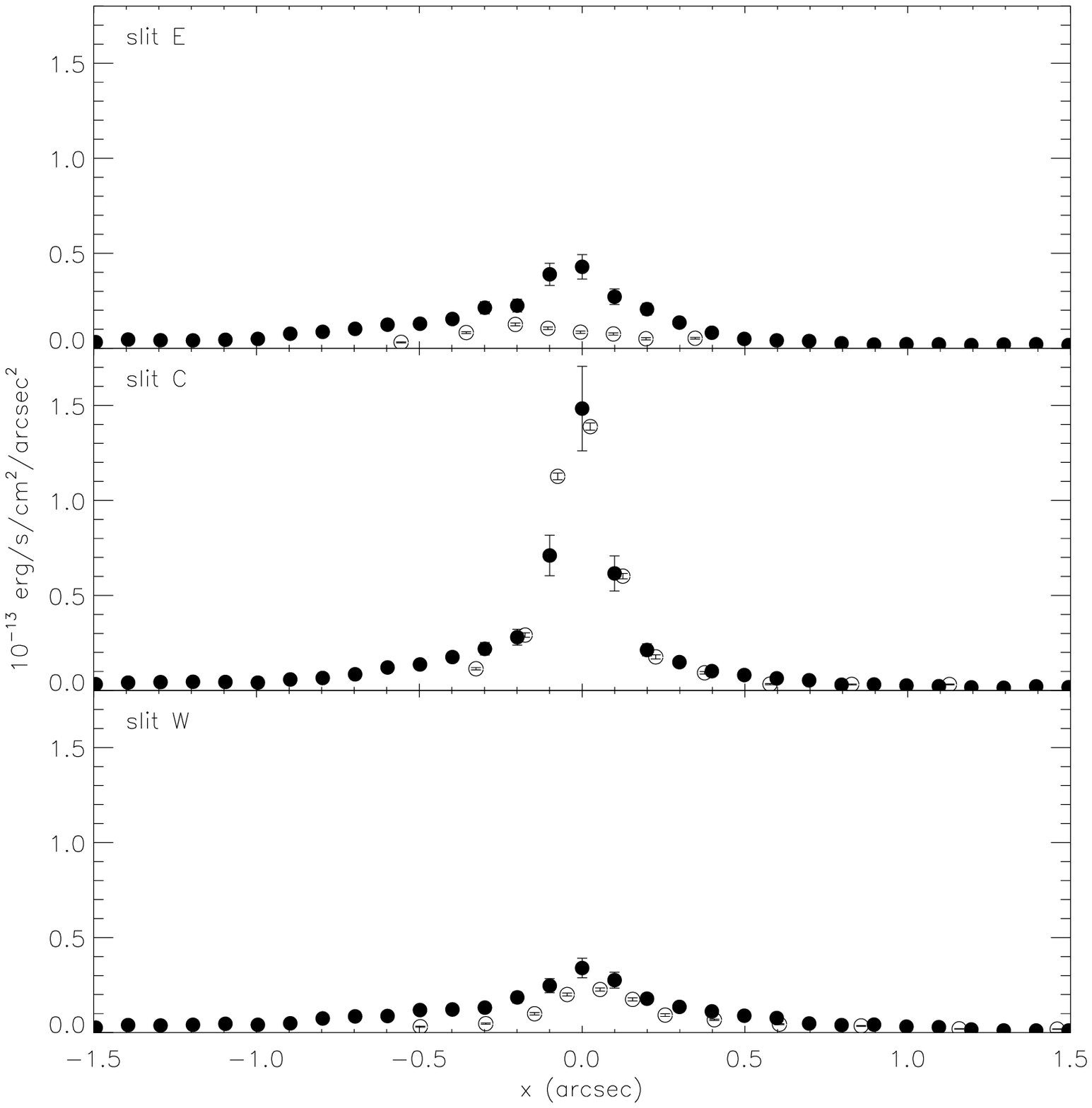}}
\ifsubmode
\vskip3.0truecm
\addtocounter{figure}{1}
\centerline{Figure~\thefigure}
\else\figcaption{\figcapfluxcompare}\fi
\end{figure}


\clearpage
\begin{figure}
\epsfxsize=0.7\hsize
\centerline{\epsfbox{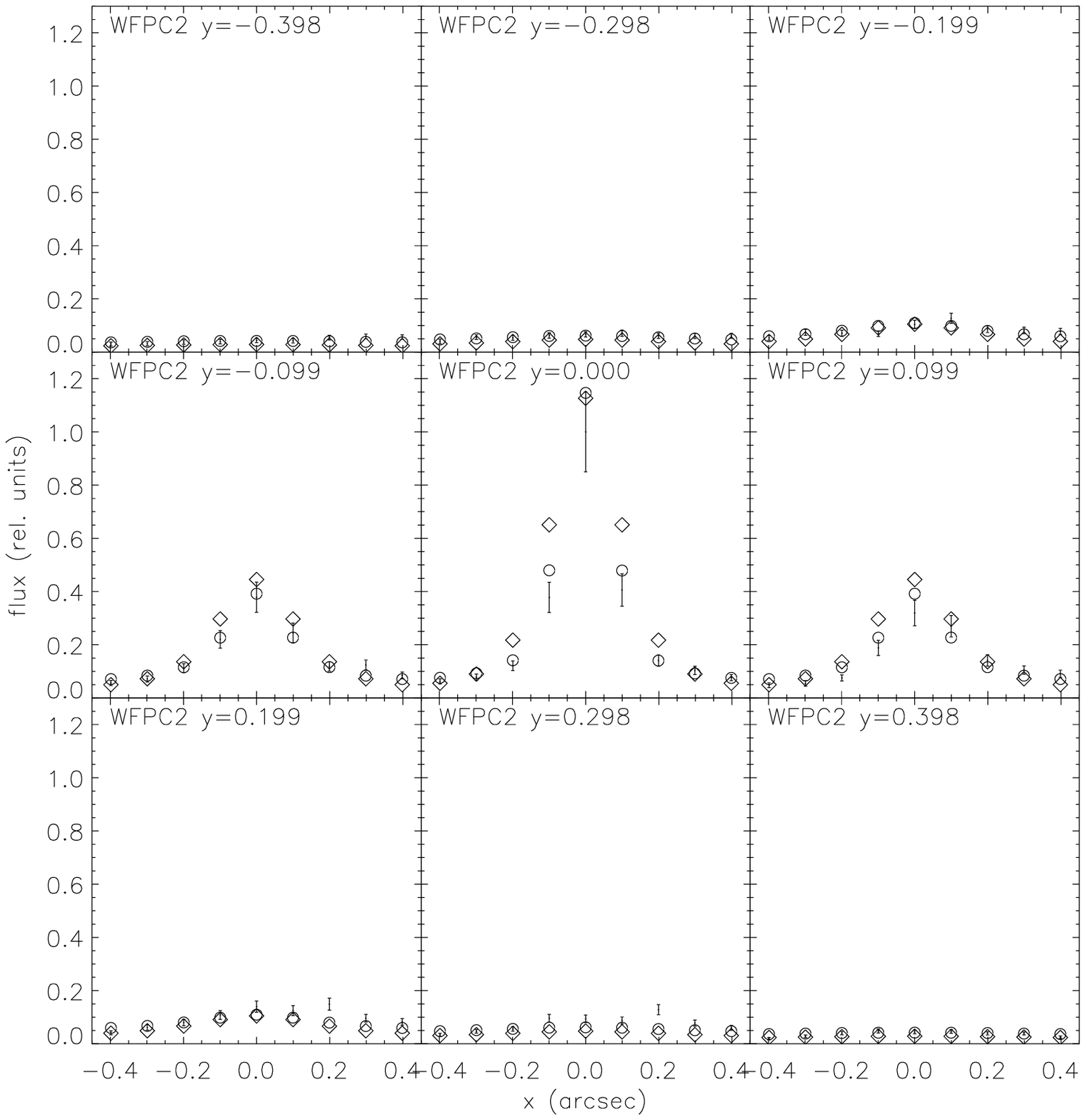}}
\epsfxsize=0.7\hsize
\centerline{\epsfbox{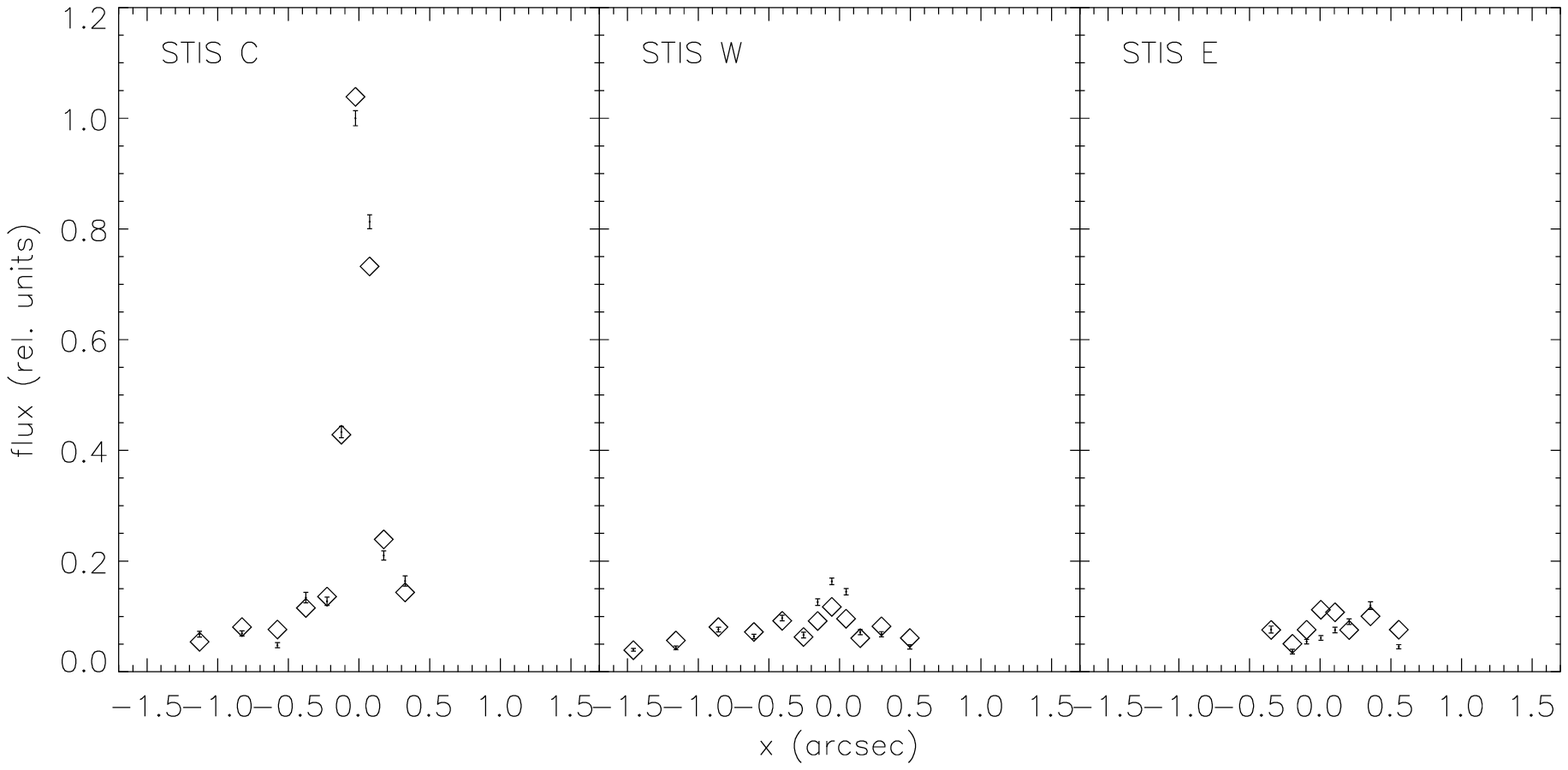}}
\epsfxsize=0.35\hsize
\centerline{\epsfbox{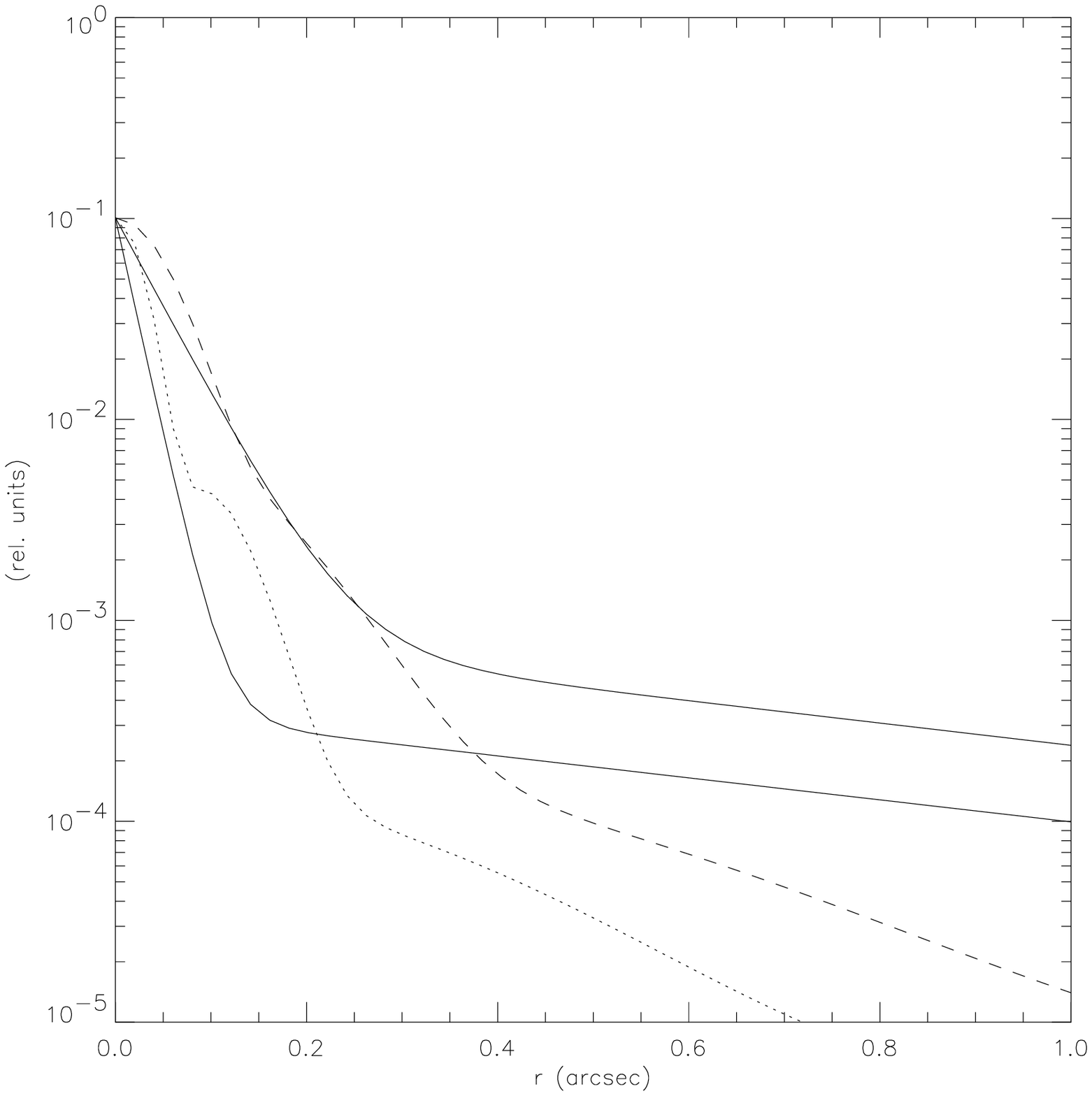}}
\ifsubmode
\vskip3.0truecm
\addtocounter{figure}{1}
\centerline{Figure~\thefigure}
\else\figcaption{\figcapfluxfit}\fi
\end{figure}


\clearpage
\begin{figure}
\centerline{\epsfbox{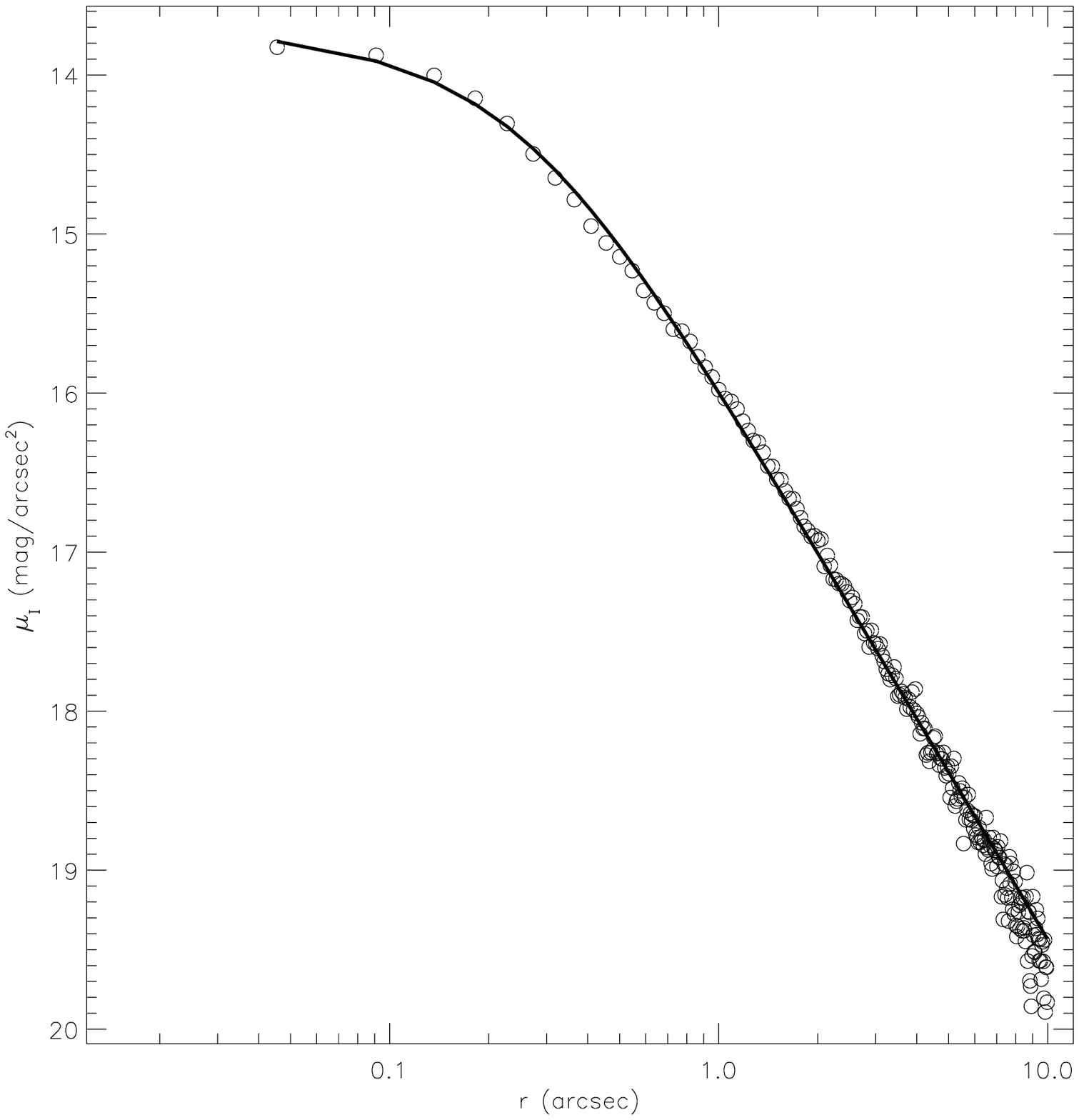}}
\ifsubmode
\vskip3.0truecm
\addtocounter{figure}{1}
\centerline{Figure~\thefigure}
\else\figcaption{\figcapsbprof}\fi
\end{figure}


\clearpage
\begin{figure}
\epsfxsize=0.9\hsize
\centerline{\epsfbox{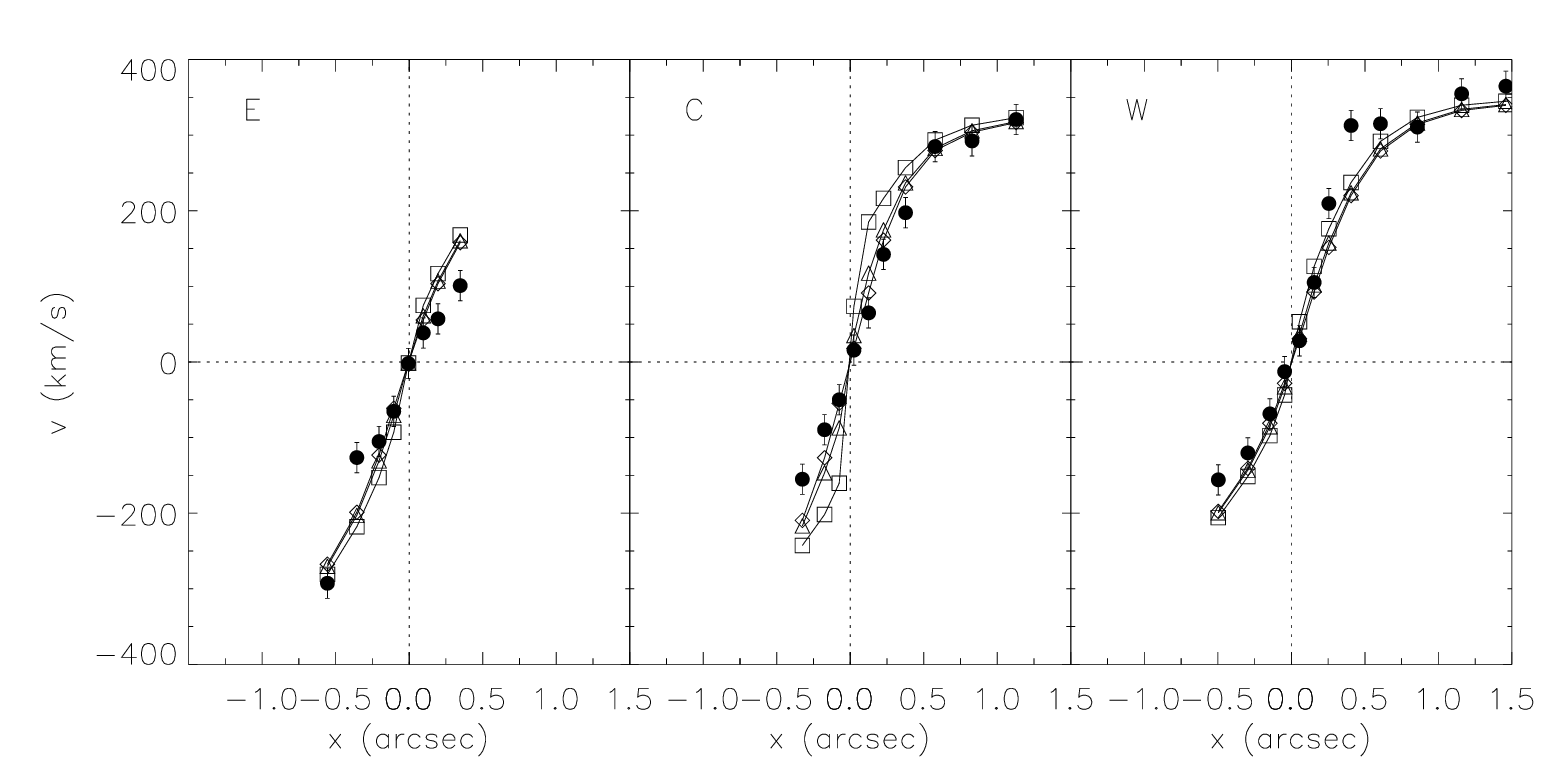}}
\epsfxsize=0.9\hsize
\centerline{\epsfbox{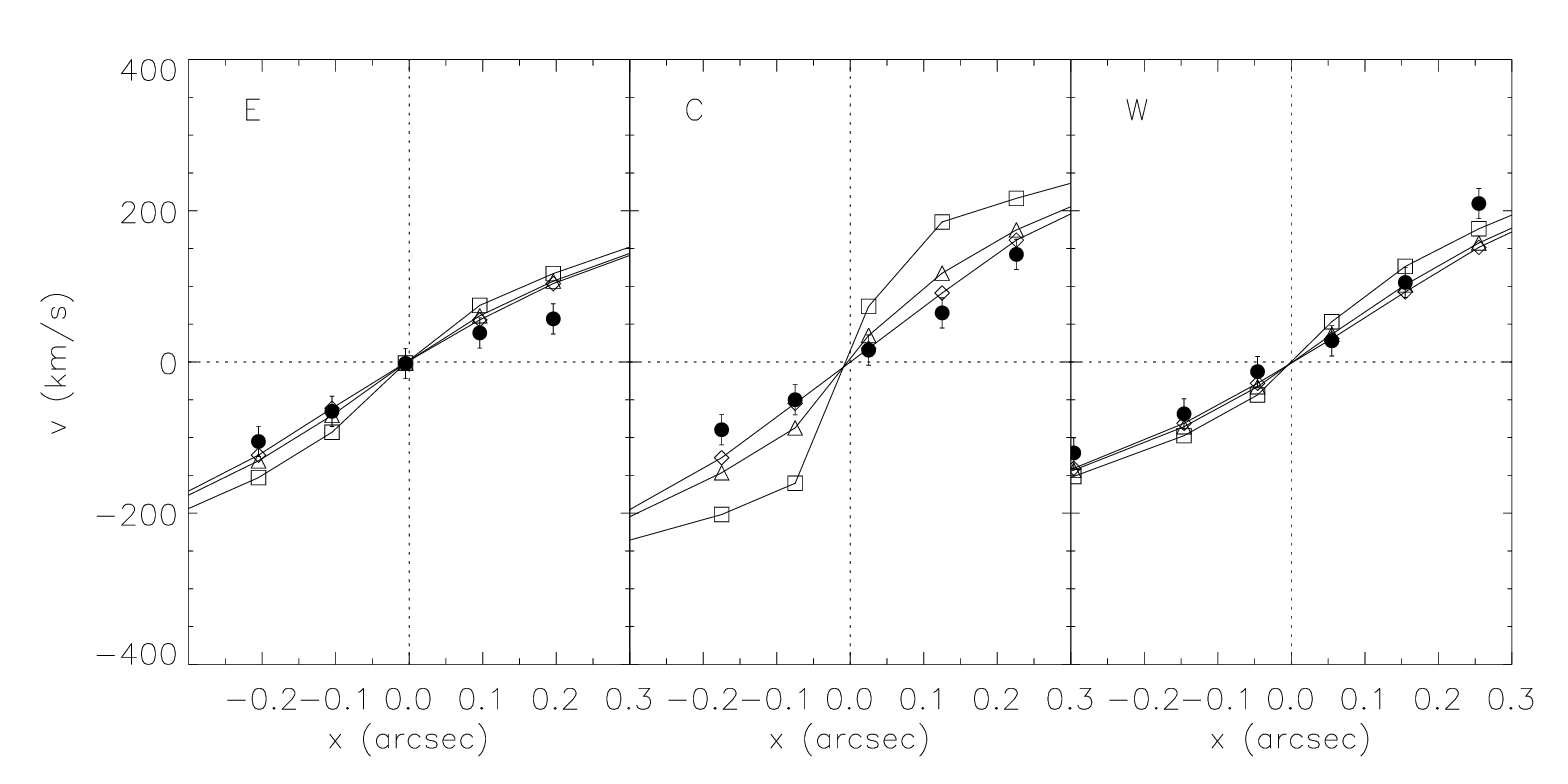}}
\ifsubmode
\vskip3.0truecm
\addtocounter{figure}{1}
\centerline{Figure~\thefigure}
\else\figcaption{\figcapmbhrange}\fi
\end{figure}


\clearpage
\begin{figure}
\epsfxsize=0.9\hsize
\centerline{\epsfbox{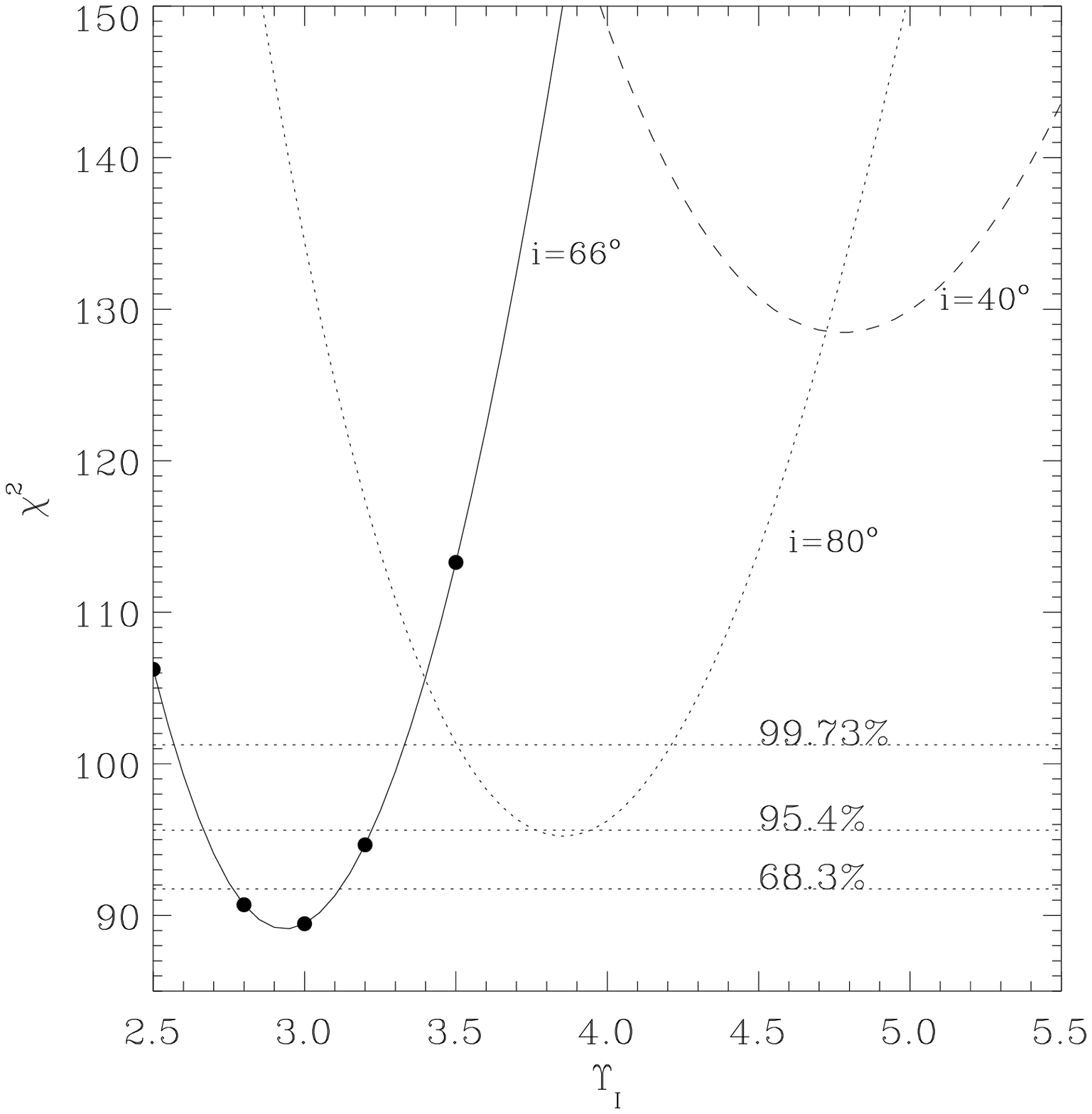}}
\ifsubmode
\vskip3.0truecm
\addtocounter{figure}{1}
\centerline{Figure~\thefigure}
\else\figcaption{\figcapupsilon}\fi
\end{figure}


\clearpage
\begin{figure}
\epsfxsize=0.9\hsize
\centerline{\epsfbox{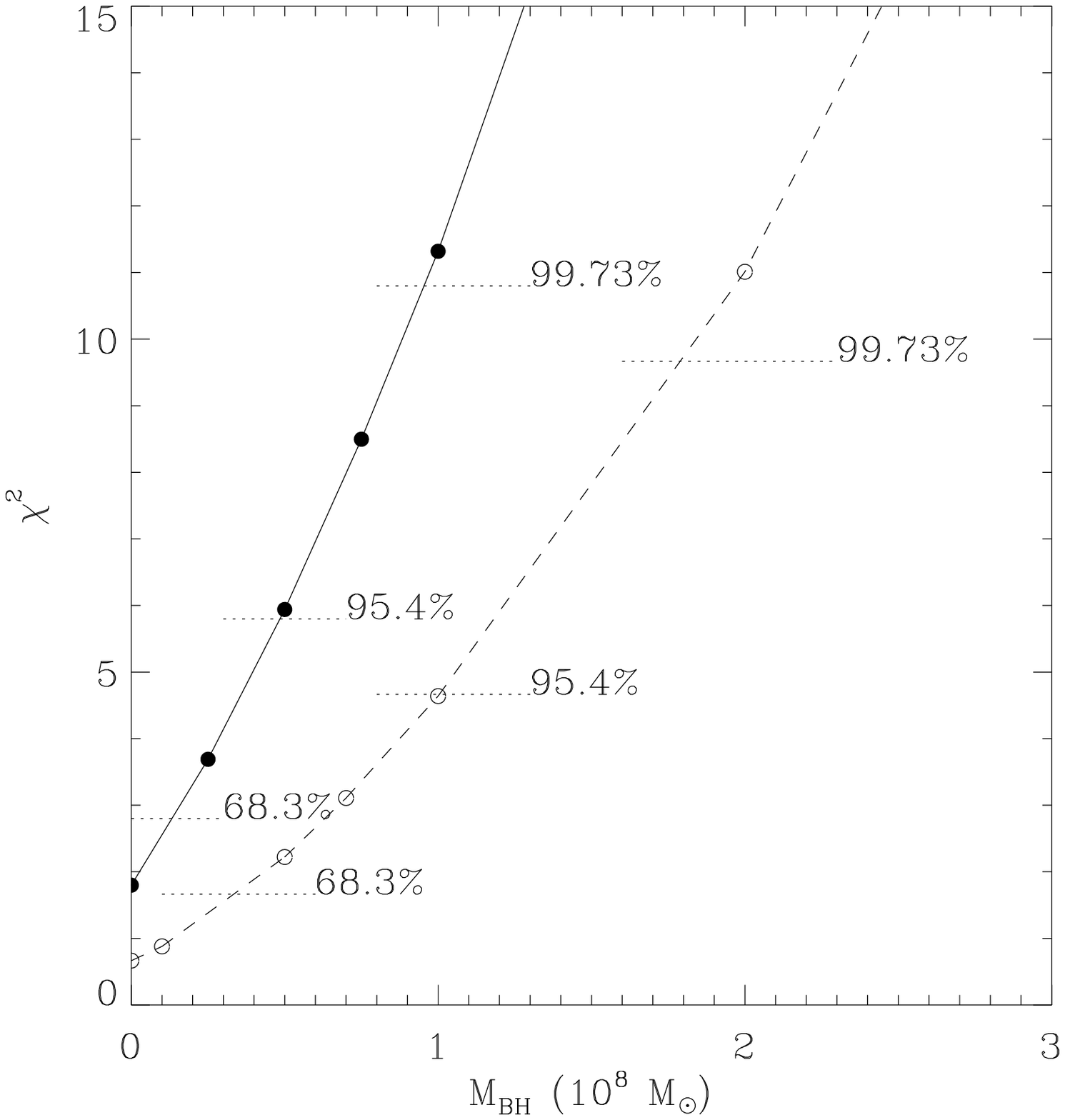}}
\ifsubmode
\vskip3.0truecm
\addtocounter{figure}{1}
\centerline{Figure~\thefigure}
\else\figcaption{\figcapchisquare}\fi
\end{figure}


\clearpage
\begin{figure}
\epsfxsize=0.9\hsize
\centerline{\epsfbox{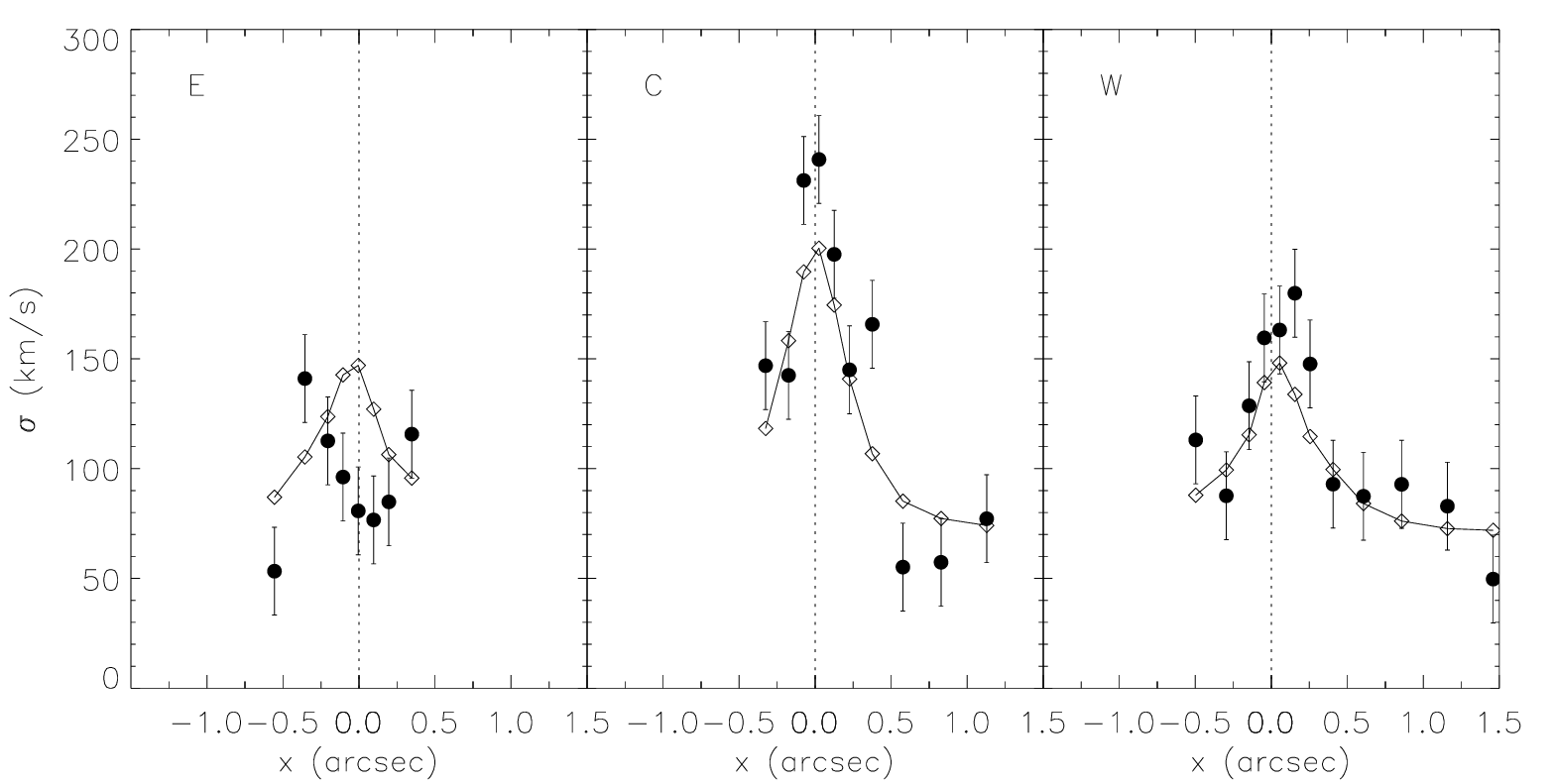}}
\ifsubmode
\vskip3.0truecm
\addtocounter{figure}{1}
\centerline{Figure~\thefigure}
\else\figcaption{\figcapturb}\fi
\end{figure}


\clearpage
\begin{figure}
\epsfxsize=0.9\hsize
\centerline{\epsfbox{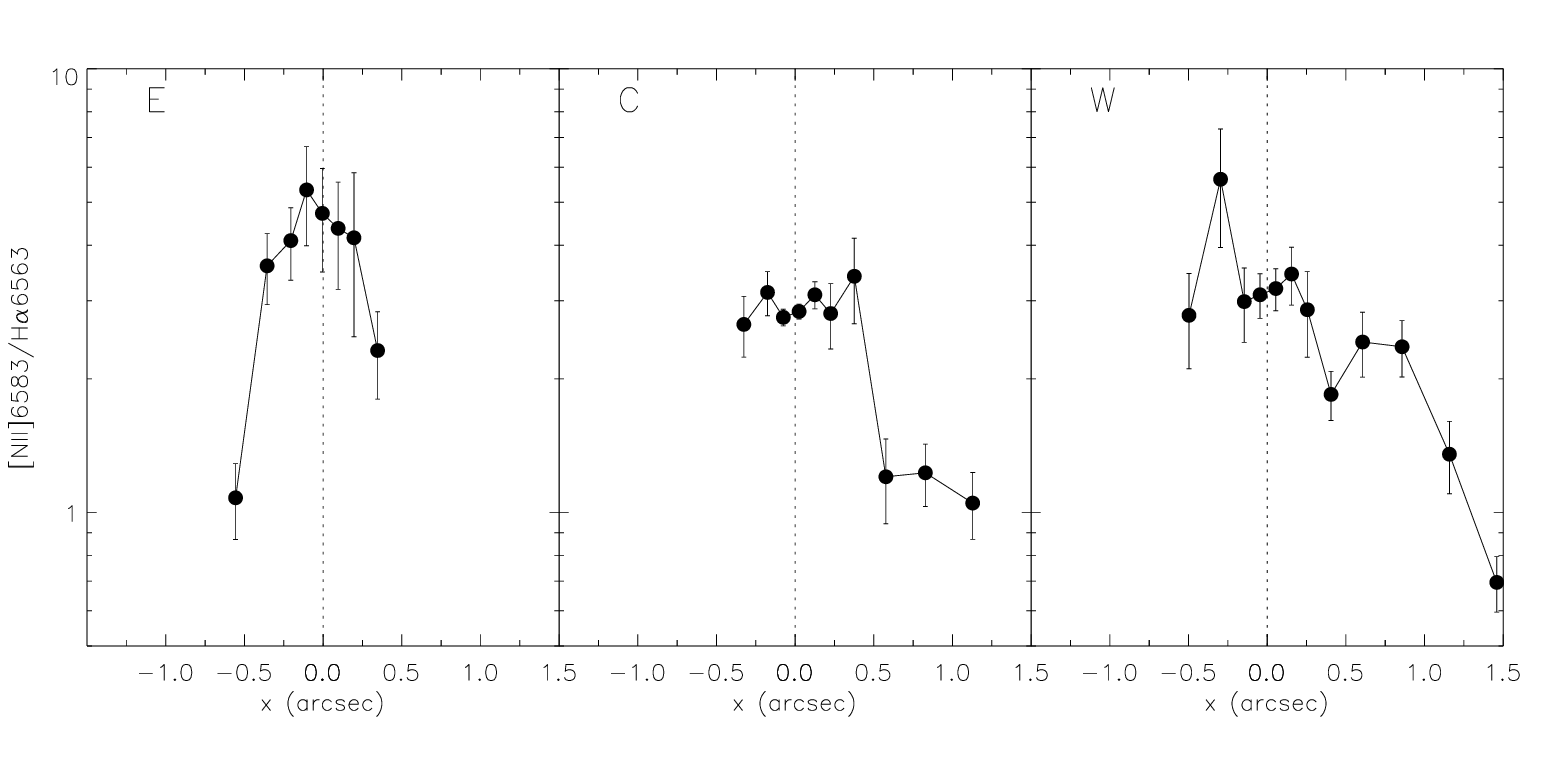}}
\ifsubmode
\vskip3.0truecm
\addtocounter{figure}{1}
\centerline{Figure~\thefigure}
\else\figcaption{\figcaplineratio}\fi
\end{figure}


\clearpage
\begin{figure}
\epsfxsize=0.9\hsize
\centerline{\epsfbox{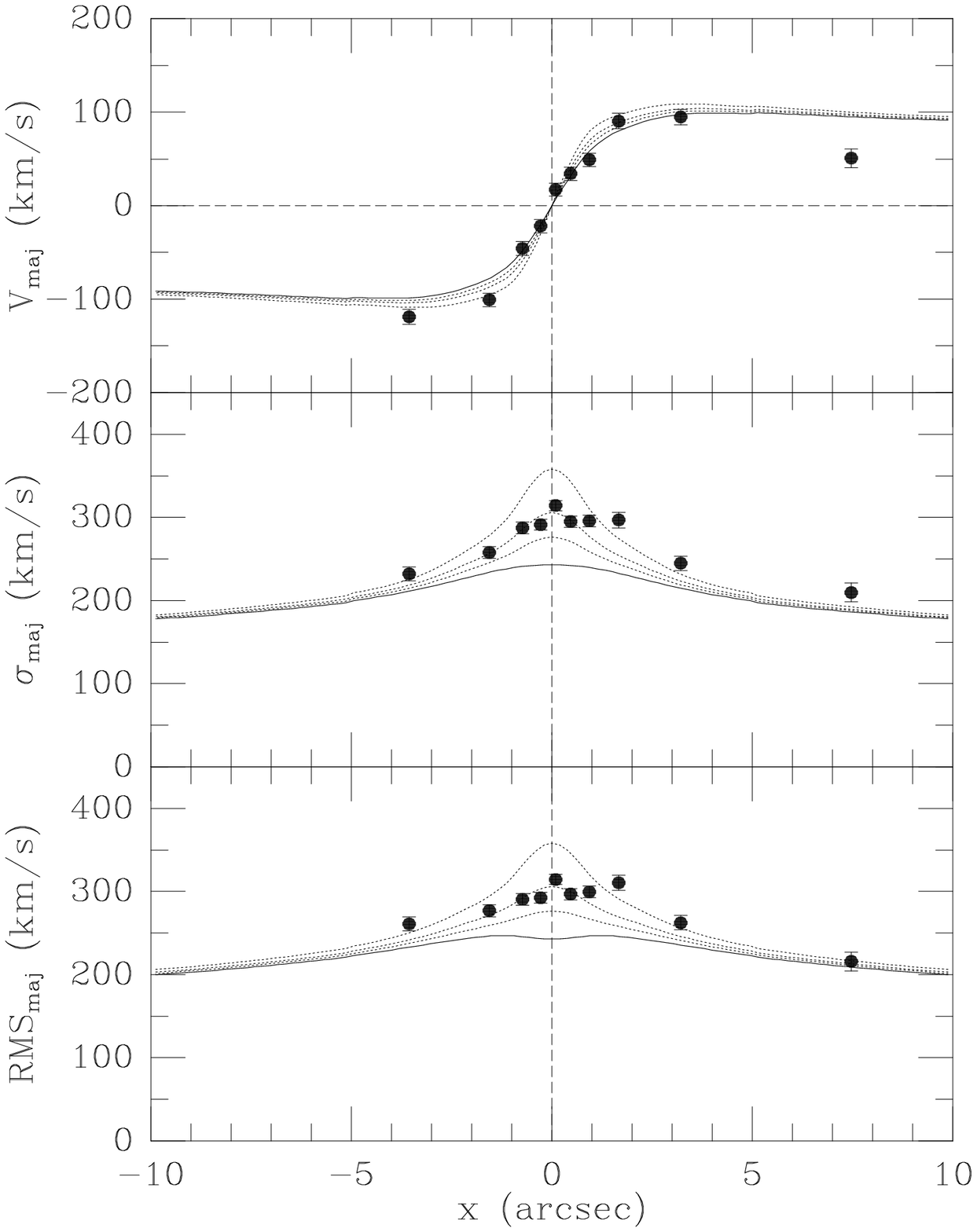}}
\ifsubmode
\vskip3.0truecm
\addtocounter{figure}{1}
\centerline{Figure~\thefigure}
\else\figcaption{\figcapstarkin}\fi
\end{figure}


\clearpage
\begin{figure}
\epsfxsize=1.0\hsize
\centerline{\epsfbox{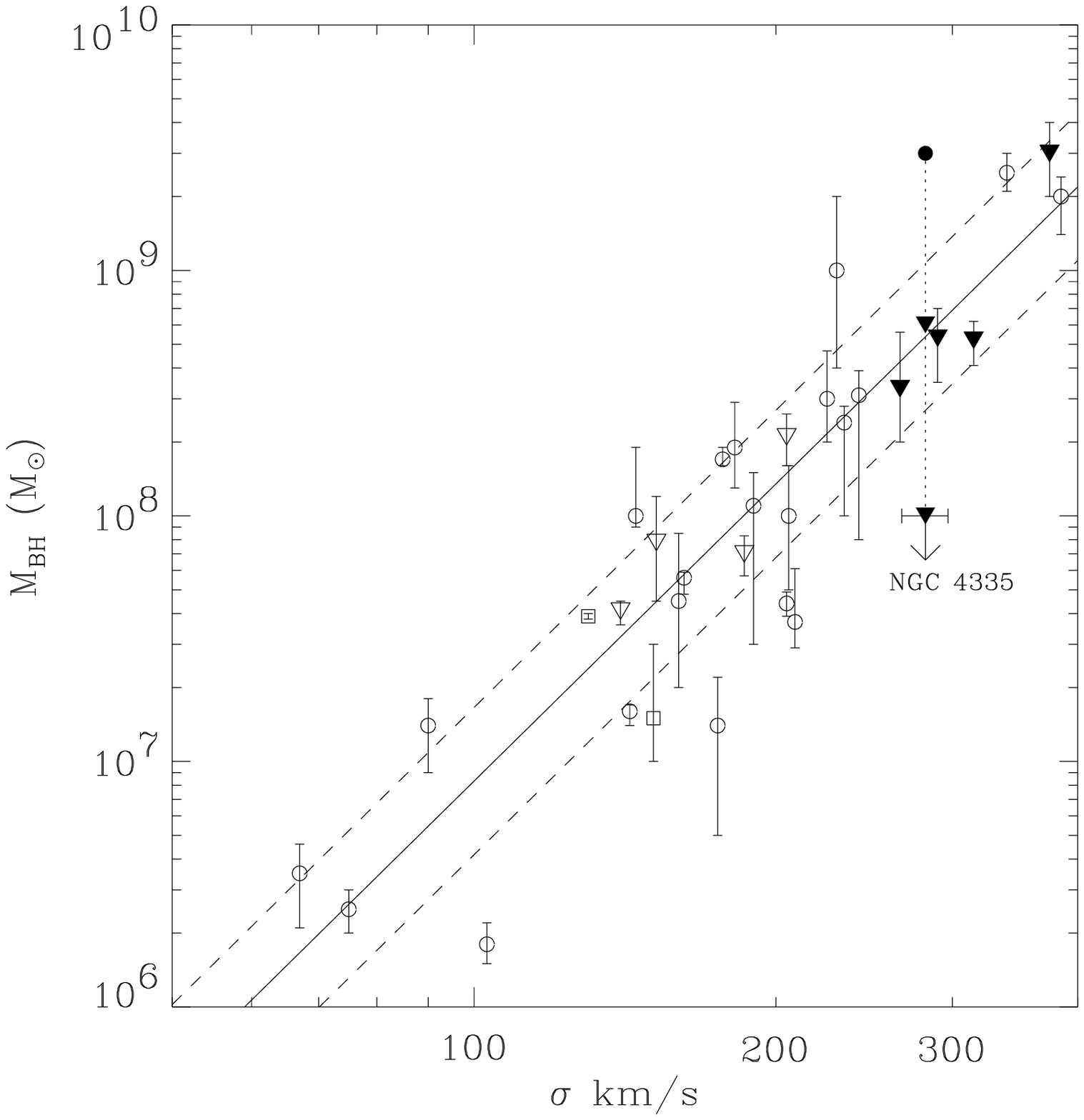}}
\ifsubmode
\vskip3.0truecm
\addtocounter{figure}{1}
\centerline{Figure~\thefigure}
\else\figcaption{\figcapbhsigma}\fi
\end{figure}


\fi


\clearpage
\ifsubmode\pagestyle{empty}\fi


\begin{deluxetable}{lrrr}
\tablewidth{\hsize}
\renewcommand{\tabcolsep}{46pt}
\tablecaption{HST/WFPC2 images: observational setup\label{t:WFPC2}}
\tablehead{
\colhead{filter} & \colhead{$\lambda_0$} & \colhead{$\Delta \lambda$} &
                   \colhead{$T_{\rm exp}$} \\ & \colhead {(\AA)} &
                 \colhead{(\AA)} & \colhead{(s)} \\
\colhead{(1)} & \colhead{(2)} & \colhead{(3)} & \colhead{(4)} \\
}
\startdata
F555W    & 5473 & 1225 & 460 \\
F814W    & 8000 & 1459 & 460 \\
LRF      & 6664 & 83 & 3600 \\
\enddata
\tablecomments{The filter name is listed in column~(1). Columns~(2)
and~(3) list the central wavelength of the filter and the FWHM.
Column~(4) lists the total exposure time, which for the broad-band
filters and the LRF was divided into two and three back to back
exposures, respectively.}
\end{deluxetable}


\clearpage
\begin{deluxetable}{crr}
\tablewidth{\hsize}
\renewcommand{\tabcolsep}{43pt}
\tablecaption{HST/STIS spectra: observational setup\label{t:STISsetup}}
\tablehead{
\colhead{ID} & \colhead{$y$} &
               \colhead{$T_{\rm exp}$} \\
& \colhead{(arcsec)} &
               \colhead{(s)} \\
\colhead{(1)} & \colhead{(2)} & \colhead{(3)} \\
}
\startdata
E & $-0.2$ & 1560 \\
C & $ 0.0$ & 1300 \\
W & $ 0.2$ & 2154 \\
\enddata
\tablecomments{STIS spectra of NGC 4335 were obtained through three
parallel slits of $0.197''$ width. Column~(1) gives the ID for the
slit. Column~(2) lists the slit-center position perpendicular to the
major axis for each slit. Column~(3) lists the exposure times.}
\end{deluxetable}


\clearpage
\begin{deluxetable}{rrrrrrrr}
\tablewidth{\hsize}
\renewcommand{\tabcolsep}{10pt}
\tablecaption{HST/STIS spectra: {\HalphaNII} kinematics\label{t:gaskin}}
\tablehead{
  \colhead{Slit} & \colhead{Rebin} & \colhead{$x$} & \colhead{$y$} & \colhead{$V$} & \colhead{$\Delta V$} & \colhead{$\sigma$} & \colhead{$\Delta \sigma$} \\
  \colhead{} & \colhead{ } & \colhead{arcsec} & \colhead{arcsec} & \colhead{$\kms$} & \colhead{$\kms$} & \colhead{$\kms$} & \colhead{$\kms$} \\
  \colhead{(1)} & \colhead{(2)} & \colhead{(3)} & \colhead{(4)} & \colhead{(5)} & \colhead{(6)} & \colhead{(7)} & \colhead{(8)} \\
  } 
\startdata
E & 2 &  0.347 & -0.254 &  101.00 &      9.68 &    115.71 &   9.44 \\
  & 1 &  0.196 & -0.232 &   57.01 &      9.58 &     84.86 &   9.34 \\
  & 1 &  0.096 & -0.217 &   38.39 &      5.65 &     76.64 &   5.62 \\
  & 1 & -0.005 & -0.202 &   -2.00 &      5.48 &     80.75 &   5.47 \\
  & 1 & -0.105 & -0.187 &  -65.27 &      5.88 &     96.18 &   6.03 \\
  & 1 & -0.205 & -0.171 & -105.18 &      6.50 &    112.63 &   6.26 \\
  & 2 & -0.356 & -0.149 & -126.46 &      8.49 &    141.02 &   7.98 \\
  & 2 & -0.556 & -0.119 & -292.72 &      4.77 &     53.30 &   4.62 \\
\hline
C & 3 &  1.128 & -0.169 &  320.72 &      6.12 &     77.20 &   6.19 \\
  & 3 &  0.828 & -0.124 &  292.35 &      4.25 &     57.35 &   4.17 \\
  & 2 &  0.577 & -0.086 &  284.91 &      5.49 &     55.16 &   5.44 \\
  & 2 &  0.376 & -0.056 &  197.42 &     12.06 &    165.72 &  11.49 \\
  & 1 &  0.226 & -0.034 &  142.31 &      8.88 &    145.00 &   8.72 \\
  & 1 &  0.125 & -0.019 &   64.90 &      4.78 &    197.58 &   4.39 \\
  & 1 &  0.025 & -0.004 &   15.66 &      3.55 &    240.79 &   2.98 \\
  & 1 & -0.075 &  0.011 &  -49.96 &      4.06 &    231.24 &   3.32 \\
  & 1 & -0.175 &  0.026 &  -89.63 &      5.59 &    142.48 &   5.62 \\
  & 2 & -0.326 &  0.049 & -155.09 &      8.40 &    146.89 &   8.24 \\
\hline
W & 3 &  1.459 & -0.016 &  364.75 &      3.23 &     49.70 &   3.14 \\
  & 3 &  1.158 &  0.029 &  354.71 &      6.85 &     82.90 &   6.86 \\
  & 3 &  0.857 &  0.074 &  310.68 &      5.85 &     92.86 &   5.61 \\
  & 2 &  0.606 &  0.111 &  315.11 &      5.76 &     87.42 &   5.92 \\
  & 2 &  0.406 &  0.141 &  312.75 &      5.11 &     92.93 &   5.06 \\
  & 1 &  0.255 &  0.164 &  209.58 &     11.82 &    147.67 &  11.28 \\
  & 1 &  0.155 &  0.179 &  105.15 &      8.92 &    179.87 &   8.15 \\
  & 1 &  0.055 &  0.194 &   28.02 &      5.92 &    163.13 &   5.79 \\
  & 1 & -0.046 &  0.209 &  -12.79 &      6.30 &    159.55 &   6.17 \\
  & 1 & -0.146 &  0.224 &  -68.63 &      8.27 &    128.64 &   8.32 \\
  & 2 & -0.296 &  0.247 & -120.31 &      5.39 &     87.64 &   5.36 \\
  & 2 & -0.497 &  0.277 & -155.89 &      9.89 &    113.07 &   9.61 \\
\enddata 
\tablecomments{\tiny Kinematics of the {\Halpha}+[NII] emission
  lines inferred from STIS spectra of NGC 4335 through three parallel
  slits of $0.197''$ width. The kinematics observed through each slit
  are separated by a horizontal line. Column~(1) lists the slit label
  (see Section~\ref{s:spec_red} for definition).  Column~(2) lists the
  number of rebinned pixels of size $0.197''$ x $0.10142''$.
  Columns~(3)-(4) list the position of the aperture center with
  respect to the major axis (i.e., $x$-axis) and minor axis
  (i.e., $y$-axis). The uncertainties in these positions are $\sim
  0.01''$. The zero-point is at the galaxy nucleus; positive $x$ values
  lie in the direction of ${\rm PA} = 156^{\circ}$. Columns~(5)-(8)
  list the mean gas velocity $V$ and velocity dispersion $\sigma$ and
  formal random errors as determined from single Gaussian fits to the
  emission lines.}
\end{deluxetable}


\clearpage
\begin{deluxetable}{rrrrrrr}
\tablewidth{\hsize}
\renewcommand{\tabcolsep}{10pt}
\tablecaption{WHT/ISIS spectra: stellar kinematics\label{t:starkin}}
\tablehead{
\colhead{Rebin} & \colhead{$x$} & \colhead{$y$} & \colhead{$V$} & \colhead{$\Delta V$} & \colhead{$\sigma$} & \colhead{$\Delta \sigma$} \\
\colhead{ } & \colhead{arcsec} & \colhead{arcsec} & \colhead{$\kms$} & \colhead{$\kms$} & \colhead{$\kms$} & \colhead{$\kms$} \\
\colhead{(1)} & \colhead{(2)} & \colhead{(3)} & \colhead{(4)} & \colhead{(5)} & \colhead{(6)} & \colhead{(7)} \\
}
\startdata
  36 &  7.39 &   1.04 &   51 & 10 & 210 & 11 \\
  12 &  3.18 &   0.45 &   95 &  8 & 245 &  9 \\
   5 &  1.65 &   0.23 &   91 &  8 & 297 &  9 \\
   3 &  0.92 &   0.13 &   49 &  7 & 296 &  7 \\
   2 &  0.47 &   0.07 &   34 &  7 & 295 &  7 \\
   2 &  0.09 &   0.01 &   17 &  8 & 314 &  6 \\
   2 & -0.28 &  -0.04 &  -22 &  7 & 291 &  6 \\
   3 & -0.73 &  -0.10 &  -46 &  8 & 287 &  7 \\
   6 & -1.54 &  -0.22 & -101 &  7 & 258 &  7 \\
  16 & -3.52 &  -0.50 & -119 &  8 & 232 &  8 \\
\enddata
\tablecomments{Stellar kinematics inferred from a WHT/ISIS long-slit
spectrum of NGC 4335 obtained with a slit approximately along the
major axis ($1.0''$ slit width). Column~(1) lists the number of
rebinned pixels of size $1.0''$ x $0.19''$. Columns~(2)-(3) list the
position of the aperture center with respect to the the major and
minor axis (i.e., $x$/$y$-axis). The coordinate system is identical to
that used for the STIS slits in Table~\ref{t:gaskin}. Columns~(4)-(7)
list the mean velocity $V$ and velocity dispersion $\sigma$ of the
stellar absorption lines with the corresponding formal random errors
as determined from a fit to the absorption-line spectrum in pixel
space (see Section~\ref{s:stellarkin}).}
\end{deluxetable}


\clearpage
\begin{deluxetable}{rrr}
\tablewidth{\hsize}
\renewcommand{\tabcolsep}{10pt}
\tablecaption{PSF STIS\label{t:psfstis}}
\tablehead{
  \colhead{$i$} & \colhead{$\gamma$} & \colhead{$\sigma$} \\
  \colhead{ } & \colhead{} & \colhead{(arcsec)} \\
  \colhead{(1)} & \colhead{(2)} & \colhead{(3)} \\
  } \startdata
1 &  0.674510 &  0.026922 \\
2 & -0.606168 &  0.045302 \\
3 &  0.798732 &  0.066942 \\
4 &  0.084999 &  0.272356 \\
5 &  0.047927 &  0.860743 \\
\enddata \tablecomments{The parameters of the five Gaussians which
  together represent the STIS PSF as PSF($r$)=$\sum_{i=1}^{5}
  \frac{\gamma_i}{2\pi\sigma_i} e^{-1/2 (r/\sigma_i)^2}$. The
  parameters were obtained through a fit to the STIS PSF produced with
  the Tiny Tim software (Krist \& Hook 2001). Differences between the
  actual PSF and the analytical fit cause differences in the modeled
  kinematics which are smaller than the errors in the data.}
\end{deluxetable}




\begin{thebibliography}{}

 
\bibitem[]{Bar01} Barth A.J., Sarzi M.R., Rix H.-W., Ho L.C., Filippenko A.V., Sargent
W.L.W., 2001, ApJ, 555, 685

\bibitem[]{Bin87} Binney J.J., Tremaine S.D., 1987, Galactic Dynamics,
Princeton University Press

\bibitem[]{Bir96} Biretta J., et al., 1996, Wide Field and Planetary Camera 2 Instrument
Handbook, Version 4.0 (Baltimore: Space Telescope Science Institute)

\bibitem[]{Boh78} Bohlin R.C., Savage B.D., Drake F.J., 1978, ApJ, 224, 132
  
\bibitem[]{Cap02} Cappellari M., \etal, 2002, \apj, in press [astro-ph/0202155]

\bibitem[]{deB96} de Bruijne J.H.J., van der Marel R.P., de Zeeuw P.T., 1996, MNRAS, 282, 909

\bibitem[]{deZ01} de Zeeuw P.T., 2001, in ESO Conference on Black Holes in Binaries
and Galactic Nuclei, eds.\ L.\ Kaper, E.P.J.\ van den Heuvel, 78

\bibitem[]{deV91} de Vaucouleurs G., de Vaucouleurs A., Corwin H.G. Jr., Buta R.J.,
Paturel G., Fouqu\'{e} P., 1991, Third Reference Catalogue of Bright
Galaxies, Springer-Verlag (RC3)

\bibitem[]{Dia01} Diaz-Miller R.I., Goudfrooij P., 2001, STIS Instrument Science Report 2000-04

\bibitem[]{Dop97} Dopita M.A., Koratkar A.P., Allen M.G., Tsvetanov Z.I., Ford H.C.,
Bicknell G.V., Sutherland R.S., 1997, ApJ, 490, 202

\bibitem[]{Fab76} Faber S.M., Jackson R.E., 1976, ApJ, 204, 668

\bibitem[]{Fan74} Fanaroff B.L., Riley F.M., 1974, MNRAS, 167, 31

\bibitem[]{Fer02} Ferrarese L.C., in Current High-Energy Emission around Black Holes, ed.\ C.-H.\ Lee (Singapore: World Scientific), in press [astro-ph/0203047]

\bibitem[]{Fer96} Ferrarese L.C., Ford H.C., Jaffe W., 1996, ApJ, 470, 444

\bibitem[]{Fer99} Ferrarese L.C., Ford H.C., 1999, ApJ, 515, 583

\bibitem[]{Fer00} Ferrarese L.C., Merritt D., 2000, ApJ, 539, L9

\bibitem[]{Fra98} Franceschini A., Vercellone S., Fabian A.C., 1998, MNRAS, 297, 817

\bibitem[]{Geb00} Gebhardt K., et al., 2000, ApJ, 539, L13

\bibitem[]{Geb02} Gebhardt K., et al., 2002, in preparation

\bibitem[]{Gou94} Goudfrooij P., de Jong T., 1994, A\&A, 298, 784

\bibitem[]{Gun79} Gunn J.E., 1979, in Active galactic nuclei (Cambridge: Cambridge University Press), 213

\bibitem[]{Har94} Harms R.J., et al., 1994, ApJ, 435, L35

\bibitem[]{Ho99} Ho L.C., 1999, in Observational Evidence for Black Holes in the
Universe, ed.\ S.K.\ Chakrabarti (Dordrecht: Kluwer), 157

\bibitem[]{Huc83} Huchra J., Davis M., Latham D., Tonry J., 1983, ApJS, 52, 89

\bibitem[]{Kim98} Kimble R., et al., 1998, ApJl, 492, 83

\bibitem[]{Kna89} Knapp G.R., Guhathakurta P., Kim D.-W., Jura M.A., 1989, ApJS, 70, 329

\bibitem[]{Kor95} Kormendy J., Richstone D., 1995, ARA\&A, 33, 581

\bibitem[]{Kor02} Kormendy J., Gebhardt K., 2002, in The 20th Texas Symposium on Relativistic
Astrophysics, eds.\ H.\ Martel, J.C.\ Wheeler, AIP, in press,
[astro-ph/0105230]

\bibitem[]{Kri01} Krist J., Hook R., 2001, {\tt http://www.stsci.edu/software/tinytim/}

\bibitem[]{Led01} Ledlow M.J., Owen F.N., Yun M.S., Hill J.M., 2001, ApJ, 552, 120

\bibitem[]{Mac97} Macchetto F.D., Marconi A., Axon D.J., Capetti A., Sparks W.,
Crane P., 1997, ApJ 489, 579

\bibitem[]{Maci01} Maciejewski W., Binney J., 2001, \mnras, 323, 831

\bibitem[]{Mag98} Magorrian J., et al., AJ, 1998, 115, 2285

\bibitem[]{Mar99} Martel A.R., et al., 1999, ApJS, 122, 81

\bibitem[]{Men83} Mendoza C., 1983, Planetary Nebualae (IAU Symposium No.~103)
ed.\ D.R.\ Flower (Dordrecht:Reidel), 143

\bibitem[]{Noe02} Noel-Storr, et al., 2002, in The central kpc of starbursts and AGN: the
La Palma connection, eds.\ J.H.\ Knapen, J.E.\ Beckman, I.\ Shlosman,
T.J.\ Mahoney (San Francisco: ASP), 249, 375 

\bibitem[]{Pat97} Paturel G., et al., 1997, A\&AS, 124, 109
  
\bibitem[]{Pre92} Press W.H., Teukolsky S.A., Vetterling W.T., Flannery B.P.,
  1992, Numerical Recipes (Cambridge: Cambridge University Press)

\bibitem[]{Ric98} Richstone D., et al., 1998, Nature, 395A, 14


\bibitem[]{Sar01} Sarzi M., et al., 2001, ApJ, 550, 65

\bibitem[]{Sch98} Schlegel D.J., Finkbeiner D.P., Davis M., 1998, ApJ, 500, 525

\bibitem[]{Tra94} Trauger J.T., 1994, ApJL, 435, 3L

\bibitem[]{Tra01} Tran H.D., Tsvetanov Z., Ford H.C., Davies J., Jaffe W., van 
den Bosch F.C., Rest A., 2001, AJ, 121, 2928

\bibitem[]{Tre02} Tremaine S., \etal, 2002, ApJ in press, [astro-ph/0203468]

\bibitem[]{vdM91} van der Marel R.P., 1991, MNRAS, 253, 710

\bibitem[]{vdM93} van der Marel R.P., Franx M., 1993, ApJ, 407, 525

\bibitem[]{vdM94a} van der Marel R.P., 1994, MNRAS, 270, 271
  
\bibitem[]{vdM94b} van der Marel R.P., Evans N.W., Rix H.-W., White S.D.M., de
  Zeeuw P.T., 1994, MNRAS, 271, 99

\bibitem[]{vdM97} van der Marel R.P., 1997, in The 1997 HST Calibration Workshop, eds.\ S.~Casertano et al.\ (Baltimore: Space Telescope Science Institute),
443

\bibitem[]{vdM98} van der Marel R.P., van den Bosch F.C., 1998, AJ, 116, 2220

\bibitem[]{vdM99} van der Marel R.P., 1999, in Galaxy Interactions at Low and High
Redshift, eds.\ J.E.\ Barnes, D.B.\ Sanders (Dordrecht:Kluwer Academic Publishers), 333

\bibitem[]{Ver99} Verdoes Kleijn G.A., Baum S.A., de Zeeuw P.T., O'Dea C.P., 1999,
  AJ, 118, 2592
  
\bibitem[]{Ver00} Verdoes Kleijn G.A., van der Marel R.P., Carollo C.M., de Zeeuw
  P.T., 2000, AJ, 120, 1221

\bibitem[]{Wad02} Wada K., Meurer G., Norman C.A., 2002, ApJ, in press.

\bibitem[]{Wro02} Wrobel J.M., Machalski J., Condon J.J., 2002, in preparation

\bibitem[]{Xu00} Xu C., Baum S.A., O'Dea C.P., Wrobel J.M., Condon J.J., 2000,
AJ, 120, 2950
\end{thebibliography}
\end{document}